\documentclass[fleqn,usenatbib]{mnras}

\usepackage[T1]{fontenc}
\usepackage{ae,aecompl}
 
\usepackage{graphicx}	
\usepackage{amsmath}	

\usepackage{amssymb}	
\usepackage{float}
\usepackage{caption}
\usepackage{subcaption}

\title[Velocity structure in the Ophiuchus cluster]{Measuring the ICM velocity structure in the Ophiuchus cluster} 

\author[Gatuzz et al.]{
Efrain Gatuzz$^{1}$\thanks{E-mail: egatuzz@mpe.mpg.de},
J. S. Sanders$^{1}$,
K. Dennerl$^{1}$,
A. Liu$^{1}$,
A. C. Fabian$^{2}$,
C. Pinto$^{3}$,\newauthor
D. Eckert$^{4}$
H. Russell$^{5}$,
T. Tamura$^{6}$,
S. A. Walker$^{7}$
and J. ZuHone$^{8}$
\\
$^{1}$ Max-Planck-Institut f\"ur extraterrestrische Physik, Gie{\ss}enbachstra{\ss}e 1, 85748 Garching, Germany\\
$^{2}$ Institute of Astronomy, Madingley Road, Cambridge CB3 0HA, UK\\ 
$^{3}$ INAF - IASF Palermo, Via U. La Malfa 153, I-90146 Palermo, Italy \\
$^{4}$ Department of Astronomy, University of Geneva, Ch. d\rq Ecogia 16, CH-1290 Versoix, Switzerland \\
$^{5}$ School of Physics \& Astronomy, University of Nottingham, University Park, Nottingham NG7 2RD, UK\\ 
$^{6}$ Institute of Space and Astronautical Science (ISAS), Japan Aerospace Exploration Agency (JAXA) Kanagawa 252-5210, Japan\\  
$^{7}$ Department of Physics and Astronomy, University of Alabama in Huntsville, Huntsville, AL 35899, USA\\
$^{8}$ Harvard-Smithsonian Center for Astrophysics, 60 Garden Street, Cambridge, MA, 02138, USA
}

\date{Accepted XXX. Received YYY; in original form ZZZ} 
\pubyear{2022} 
\begin{document}
 \label{firstpage}
\pagerange{\pageref{firstpage}--\pageref{lastpage}}
\maketitle 
\begin{abstract}
We have found evidence of bulk velocities following active galactic nucleus (AGN) bubbles in the Virgo cluster and galaxy motions in the Centaurus cluster. In order to increase the sample and improve our understanding of the intracluster medium (ICM), we present the results of a detailed mapping of the Ophiuchus cluster with {\it XMM-Newton} to measure bulk flows through very accurate Fe~K measurements. To measure the gas velocities we use a novel EPIC-pn energy scale calibration, which uses the Cu K$\alpha$ instrumental line as reference for the line emission. We created 2D spectral maps for the velocity, metallicity, temperature, density, entropy and pressure with a spatial resolution of 0.25$'$ ($\sim 26$~kpc). The ICM velocities in the central regions where AGN feedback is most important are similar to the velocity of the brightest cluster galaxy (BCG). We have found a large interface region where the velocity changes abruptly from blueshifted to redshifted gas which follows a sharp surface brightness discontinuity. We also found that the metallicities and temperatures do not change as we move outwards from the giant radio fossil previously identified in radio observations of the cluster. Finally, we have found a contribution from the kinetic component of $<25\%$to the total energy budget for large distances.
\end{abstract} 

\begin{keywords}
X-rays: galaxies: clusters -- galaxies: clusters: general -- galaxies: clusters: intracluster medium -- galaxies: clusters: individual: Ophiuchus
\end{keywords} 

\section{Introduction}\label{sec_in} 

Studying the kinematics of the ICM is important to fully understand the physical properties of such environment. Measurement of velocities can help to constrain AGN feedback models, since the energy distribution from AGN feedback within the bulk of the cluster depends on the balance between sound waves or shocks and turbulence \citep[see][for a review]{fab12b}. Turbulent motions affect cluster mass estimates through calculations of hydrostatic equilibrium because they provide additional pressure support, the non-thermal pressure provided by turbulence is about $6-10$ $\%$ \citep[e.g. ][]{lau09,eck19}. Velocity is also an excellent probe of the microphysics of the ICM, such as viscosity \citep{gas14}. Motions will also cause transport of metals in the ICM, due to uplift of metals by AGN \citep[e.g.][]{sim08,wer10}. In addition, measuring velocities should directly measure the sloshing of gas in cold fronts, which can remain for several Gyr \citep{roe12,roed13,wal18}. 
  
Simulations indicate that the ICM should contain bulk flows and random, or turbulent motions, due to the merger of other clusters and subcomponents \citep{lau09,vaz11,sch17,haj18,lim18,vaz21}. Merging substructures can generate relative bulk motions of several hundred km/s due to perturbations in the ICM \citep{asc06,ich19,vaz18,zuh18}. Furthermore, inflation of bubbles and the action of the relativistic jets by the central AGN also likely generate motions of a few hundred km/s  \citep{bru05,hei10,ran15,yan16,bam19}. Overall, there is a close connection between the velocity power spectra and the ICM physical state, identified by its temperature, density, pressure and entropy \citep{gas14}. Despite its importance, the velocity structure of the ICM remains poorly constrained observationally.

  \begin{table} 
\footnotesize
\caption{\label{tab_obsids}{\it XMM-Newton} observations of the Ophiuchus cluster.}
\centering 
\begin{tabular}{cccccccc}   
\hline
ObsID & RA & DEC & Date & Exposure \\
 &&&Start-time& (ks)\\
\hline
0105460101&17:12:36.00& -24:14:41.0&	2000-09-18&        21.3\\
0105460601&17:12:36.00& -24:14:41.0&	2001-08-30&        18.8\\
0206990701&17:09:44.89& -23:46:58.0&	2005-02-28&        18.3\\
0505150101&17:12:27.39& -23:22:19.7&	2007-09-02&        36.7\\
0505150201&17:12:27.39& -23:42:19.7&	2007-09-26&        26.7\\
0505150301&17:11:00.23& -23:22:18.1&	2008-02-20&        27.9\\
0505150401&17:12:27.39& -23:02:19.7&	2008-02-21&        27.9\\
0505150501&17:13:54.55& -23:22:18.3&	2008-02-21&        29.8\\
0862220501&17:13:14.99& -23:43:15.0&	2020-09-01&        44.9\\ 
0862220101&17:13:25.49& -23:41:15.0&	2020-08-31&        40.0\\
0862220301&17:13:32.69& -24:02:51.0&	2020-09-28&        38.5\\
0862220201&17:14:52.70& -23:43:08.0&	2021-03-18&        33.0\\ 
0880280801&17:11:57.84& -23:35:41.0&	2021-09-06&        59.9\\
0880280401&17:11:57.84& -23:35:41.0&	2021-09-17&       126.2\\
0880280901&17:11:57.84& -23:35:41.0&	2021-09-17&        118.7\\
0880280201&17:11:31.97& -23:21:05.1&	2021-09-19&       126.1\\
0880281001&17:11:31.97& -23:21:05.1&	2021-09-19&        119.8\\
0880280301&17:13:12.01& -23:21:35.4&	2021-09-21&       127.0\\
0880281101&17:13:12.01& -23:21:35.4&	2021-09-21&        11.0\\
0880280101&17:12:57.59& -23:11:15.8&	2021-09-29&       127.2\\
0880281201&17:12:57.59& -23:11:15.8&	2021-09-29&        11.0\\
0880280701&17:13:12.01& -23:21:35.4&	2022-02-26&        55.1\\
0880280601&17:11:31.97& -23:21:05.1&	2022-03-12&        54.0\\
0880281301&17:11:31.97& -23:21:05.1&	2022-03-11&         5.9\\
0880280501&17:12:57.59& -23:11:15.8&	2022-03-16&        51.0\\ 
 \\
 \hline
\end{tabular}
\end{table}

The {\it Hitomi} observatory \citep{tak16} directly measured bulk and random motions in the ICM using the Fe-K emission lines with its high spectral resolution microcalorimeter SXS X-ray detector. It measured a gradient of $150$ km/s bulk flow across $60$ kpc of the Perseus cluster core and a line-of-sight velocity dispersion of $164 \pm 10$ km/s between radii of $30-60$ kpc \citep{hit16}. These results, done for a very limited spatial region, showed that the Perseus core is not strongly turbulent, despite the obvious impact of the AGN and its jets on the surrounding ICM. Such level of turbulence may be sufficient to offset radiative cooling if driven on scales comparable with the size of the largest bubbles in the field \citep[about 20–30 kpc, see ][]{hit16}. \citet{fab17} showed that it might not be fast enough to replenish heating in the short time necessary to balance cooling, although shocks or sound waves could plausibly do this. Gravity waves, on the other hand, do not propagate efficiently radially \citep{fab17} and the production of turbulence in the presence of large-scale magnetic fields is also inefficient even when able to preserve AGN-drive bubbles \citep{bam18}. Simulation studies of Perseus-like clusters with either AGN \citep{lau17} or sloshing \citep{zuh18} show that the same level of low velocities can be achieved, even without including viscosity effects. Unfortunately, due to the loss of {\it Hitomi} we will not be able to make further measurements in other clusters or in different regions of Perseus. The next planned observatories with instruments capable of measuring velocities are likely {\sc XRISM} \citep{xri20}, to be launched in 2023 and {\sc Athena} \citep{bar18} in 2035.
 
Velocities in several systems were obtained using {\it Suzaku} by measuring the Fe-K line. \citet{tam14} placed upper limits on relative velocities of $300$ km/s over scales of $400$ kpc in Perseus. \citet{ota16} examined several clusters with {\it Suzaku}, although systematic errors from the {\it Suzaku} calibration were likely around $300$ km/s and its PSF was large. Similar attempts have also been made with Chandra. For example, \citet{liu15,liu16} studied velocity structures in several merging clusters by measuring Fe-K line using {\it Chandra} CCD data.  Low turbulence motion is also measured from line broadening and resonant scattering, limited to the cluster core, with velocities between $100-300$ km/s  \citep{san10,san13,pin15,ogo17,liu19}. Indirect measurements of velocity structure in clusters include looking at the power spectrum of density fluctuations and linking this via simulations to the velocity spectrum \citep{zhu14} or examining the magnitude of thermodynamic perturbations \citep{hof16}. However, these methods are model-dependent.

 \citet{san20} present a novel technique which consists of using background X-ray lines seen in the spectra of the {\it XMM-Newton} EPIC-pn detector to calibrate the absolute energy scale of the detector to better than $150$ km/s at Fe-K. Using this technique, \citet{san20} mapped the bulk velocity distribution of the ICM over a large fraction of the central region of  Perseus and Coma clusters. They detect evidence for sloshing associated with a cold front for the Perseus cluster. In the case of the Coma cluster, they found that the velocity of the gas is close to the optical velocities of the two central galaxies, NGC 4874 and NGC 4889, respectively. Three large {\it XMM-Newton} observation programs were approved in recent years to study the Virgo, Centaurus and Ophiuchus clusters with this method. The analysis of the Virgo cluster \citep{gat22a} shows signatures of AGN outflows (e.g. changes in the velocity for regions following the radio morphology; the lack of a gas sloshing morphology in the velocity map similar to those obtained in simulations) as well as gas sloshing (e.g. there is an overall gradient in the velocities, discontinuities in metallicity within the cold fronts). On the other hand, the study of the Centaurus cluster done by \citet{gat22b} shows that the velocity structure of the ICM is similar to the velocity structure of the main galaxies while the cold fronts are likely moving in a plane perpendicular to our line of sight with low velocity. Here, we present the analysis of the ICM velocity structure in the Ophiuchus galaxy cluster.

The Ophiuchus cluster constitutes an excellent candidate, located at z = 0.0296 \citep{dur15}, it is the second brightest galaxy cluster in the 2--10 keV sky \citep{edg90} and shows the second brightest iron line measured, after Perseus. \citep{nev09} analyzed the only {\it XMM-Newton} observation centered in the cluster together with {\it INTEGRAL} data. They modeled the $0.6-140$ keV band total emission in the central region of the cluster, concluding that the pressure of the non-thermal electrons is $\sim$1\% of that of the thermal electrons.  \citet{per09} analyzed the same observation in combination with radio data from the Very Large Array radio. They found that there is no significant radio emission, and that a two temperature model is required to fit the X-ray emission in the core, suggesting that the innermost central region of the Ophiuchus cluster is a cooling core. Analysis of {\it Chandra} observations have revealed the existence of cold fronts \citep{mil10}, suggesting substantial sloshing of the X-ray emitting gas. The cluster hosts a truncated cool core, sharply peaked and with a temperature increasing from kT $\sim$ 1 keV in the center to kT $\sim$ 9 keV at r$\sim$ 30 kpc \citep{mil10}. This core is dynamically disturbed by the cold fronts with indications for both Kelvin-Helmholtz and Rayleigh-Taylor instabilities \citep{wer16b}. The AGN itself only displays weak, point-like radio emission (i.e. there are no jets or lobes), although \citet{gia20} claim to have discovered a large cavity to the southeast of the Cluster. Such a cavity could be a very aged fossil of the most powerful AGN outburst seen in any galaxy cluster. Finally, Ophiuchus is a massive and relatively relaxed cluster \citep[however there might be a minor merger, see e.g. ][]{dur15}.

The outline of this paper is as follows. We describe the data reduction process in Section~\ref{sec_dat}. The fitting procedure and the results are shown in Section \ref{sec_fits} while a discussion of the results is included in Section~\ref{sec_dis}. Finally, Section~\ref{sec_con} presents the conclusions and summary. Throughout this paper we assume the distance of Ophiuchus to be $z=0.0296$ \citep{dur15} and a concordance $\Lambda$CDM cosmology with $\Omega_m = 0.3$, $\Omega_\Lambda = 0.7$, and $H_{0} = 70 \textrm{ km s}^{-1}\ \textrm{Mpc}^{-1} $.

\begin{figure*}
        \centering
        \begin{subfigure}{0.50\textwidth}
            \includegraphics[scale=0.22]{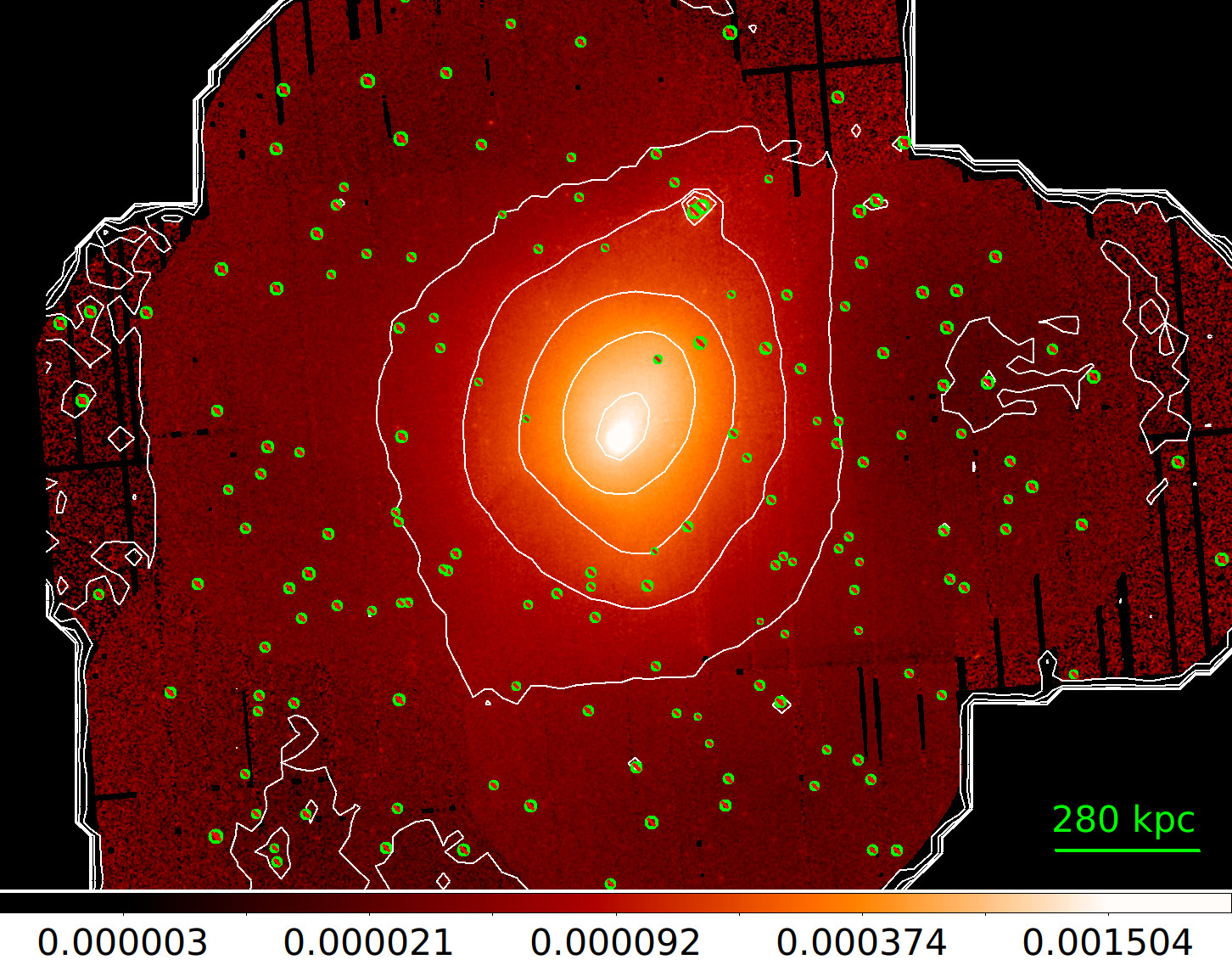}
        \end{subfigure}
          \centering
        \hspace{2.3cm}          
        \begin{subfigure}{0.32\textwidth}
        	\vspace{-0.2cm}
           \includegraphics[scale=0.107]{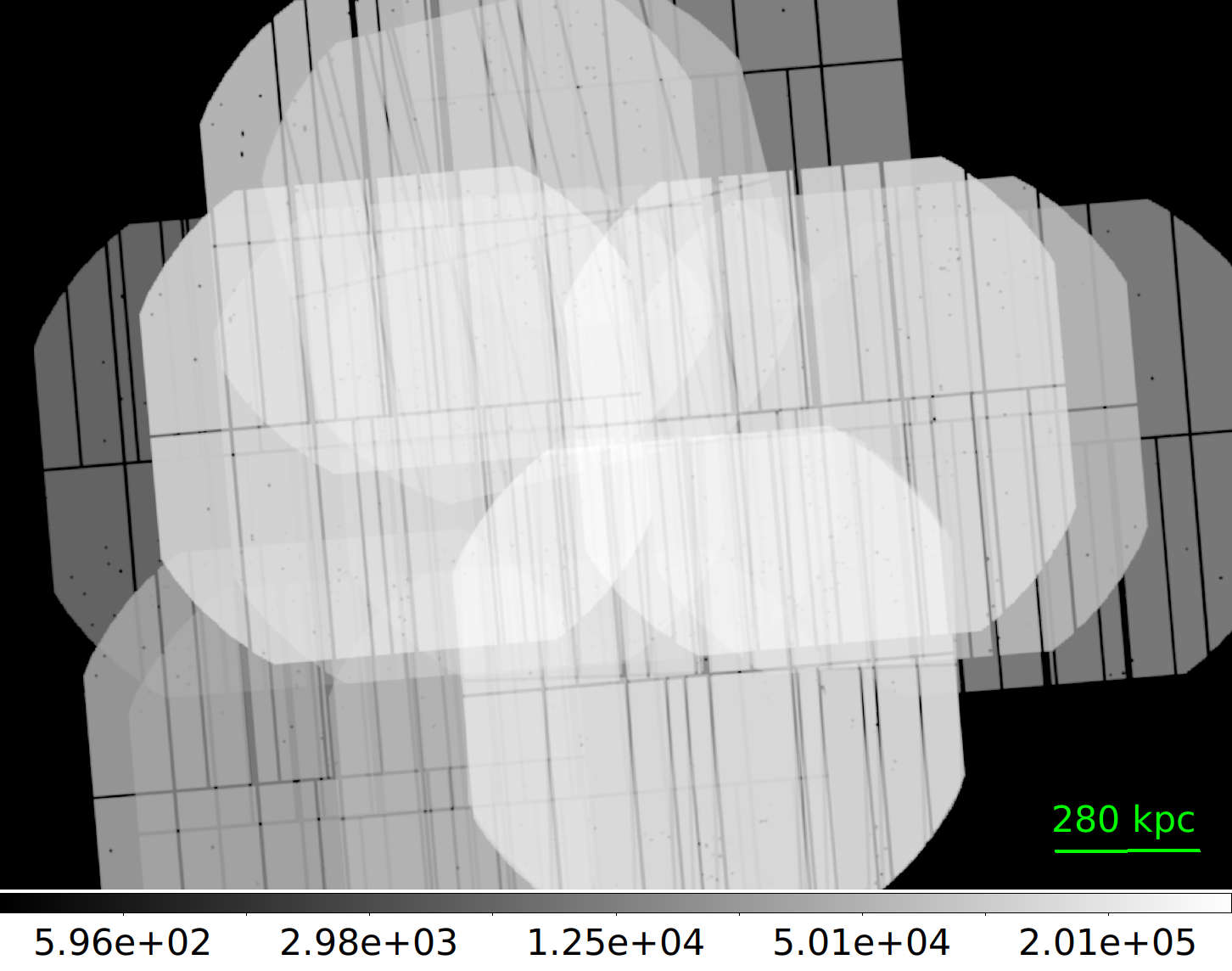}\\
           \includegraphics[scale=0.107]{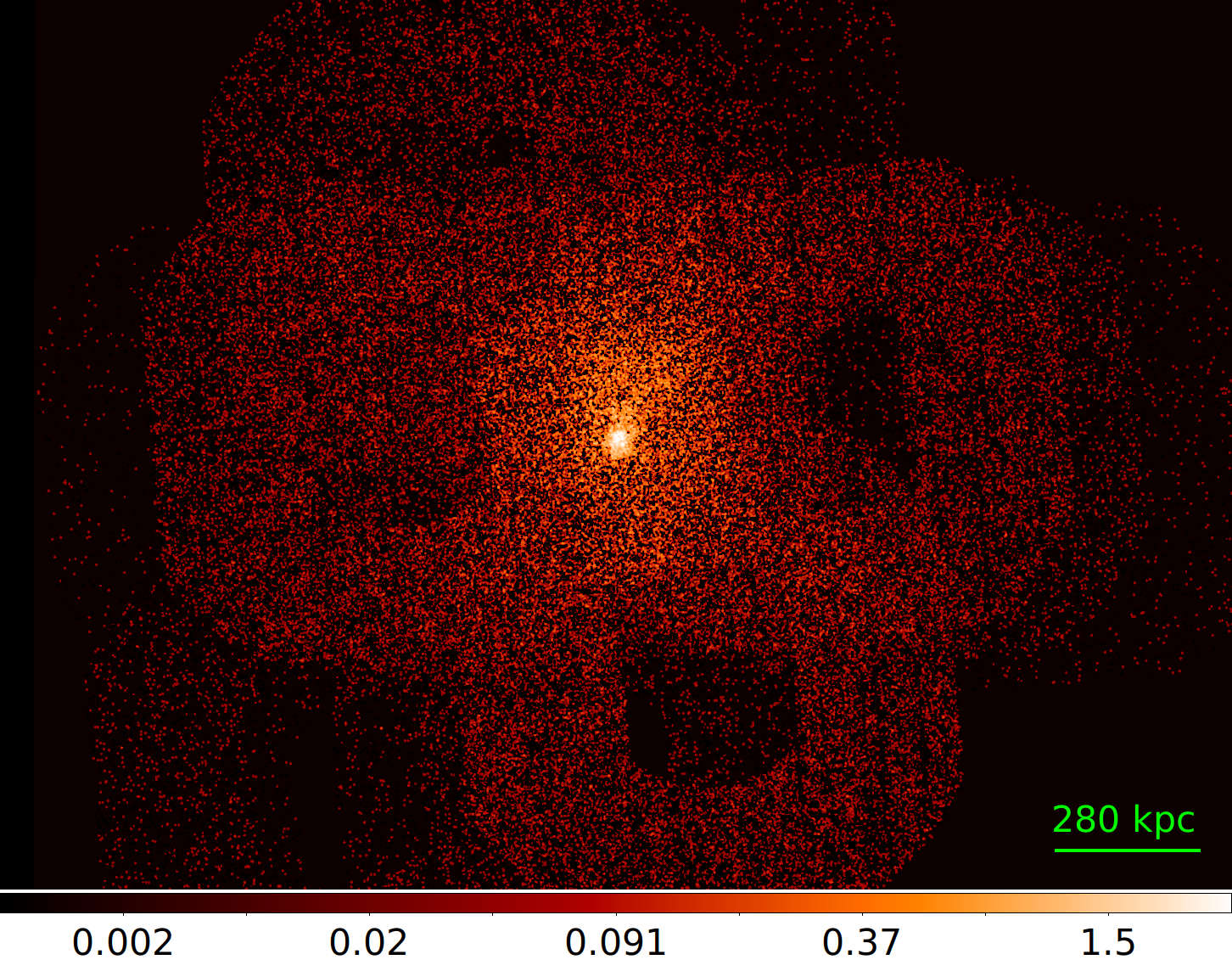}
        \end{subfigure}
        \caption{\emph{Left panel:} X-ray surface brightness exposure corrected and background subtracted of the Ophiuchus cluster in the 0.5 to 9.25 keV energy range. The point-like sources were excluded in the analysis (green circles). The 10-level contours of the X-ray image are shown in white. \emph{Top right panel:} total exposure time (s) in the 0.5-9.25 keV energy range. \emph{Bottom right panel:}  Fe-K count map, showing the number of counts in each 1.59 arcsec pixel in the Fe-K complex.} \label{fig_xray_maps} 
\end{figure*}

\begin{figure} 
\centering 
\includegraphics[width=0.485\textwidth]{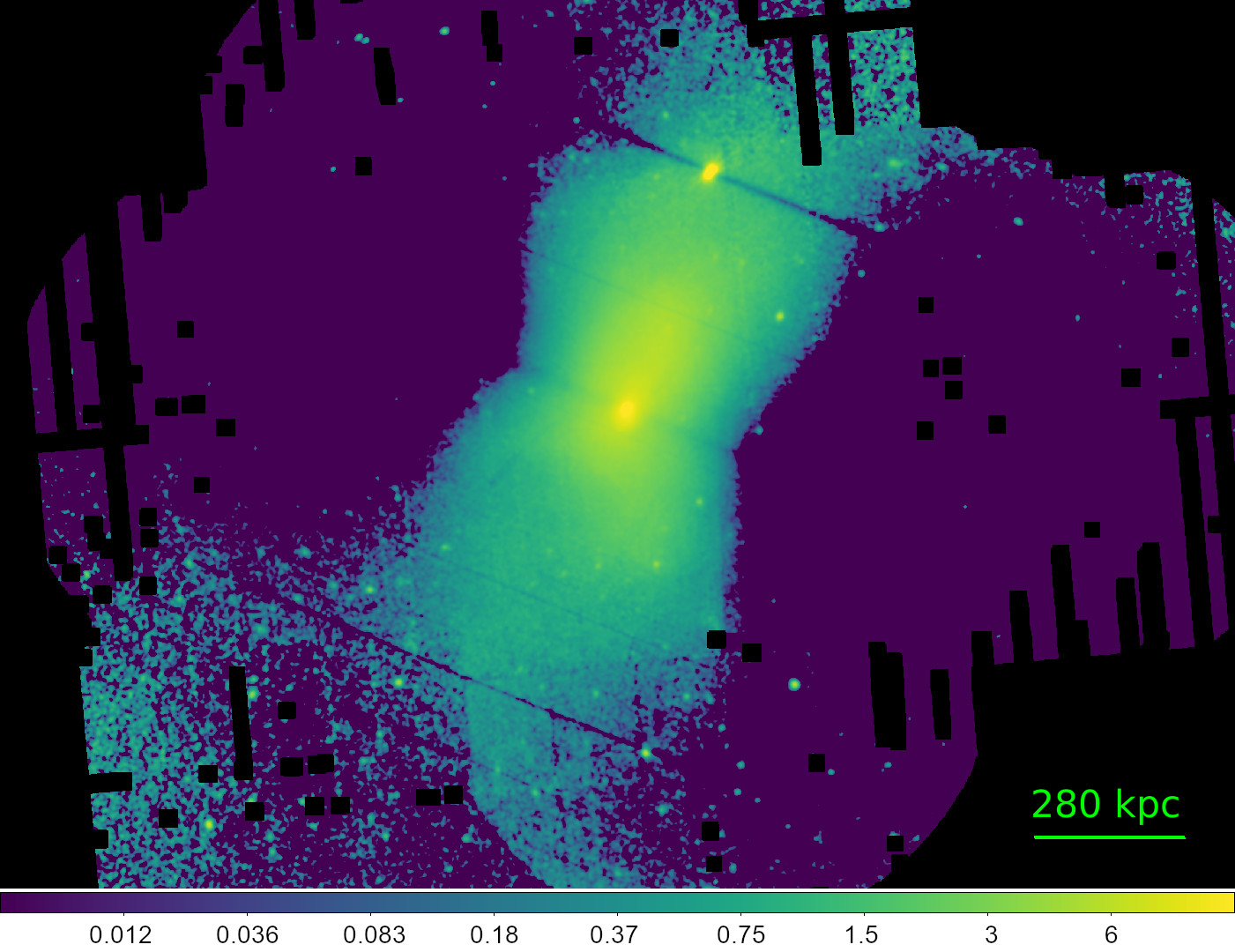}
\caption{Subtracted fractional difference from the average at each radius for the 0.5-9.25 keV X-ray surface brightness. } \label{fig_xra_sb} 
\end{figure}

\begin{table}
\scriptsize 
\caption{\label{tab_circular_fits}Ophiuchus cluster best-fit parameters for Case 1 extracted regions. }
\centering
\begin{tabular}{ccccccc}
\\
Region &\multicolumn{5}{c}{{\tt apec} model}  \\
\hline
 &$kT$ & Z& $z$   & $norm$   & cstat/dof\\ 
  &   & &  ($\times 10^{-3}$) &   ($\times 10^{-3}$)  \\  
1&$5.63_{-0.29}^{+0.22}$&$0.78\pm 0.06$&$29.14\pm 1.44$&$4.33\pm 0.06$&$1580/1532$\\
2&$8.03\pm 0.17$&$0.66\pm 0.03$&$29.83_{-1.01}^{+1.12}$&$12.57\pm 0.08$&$1707/1670$\\
3&$9.08\pm 0.21$&$0.52\pm 0.03$&$30.77_{-1.18}^{+1.11}$&$16.63\pm 0.08$&$1696/1682$\\
4&$8.69\pm 0.18$&$0.48\pm 0.02$&$29.87\pm 0.82$&$21.61\pm 0.09$&$1783/1686$\\
5&$8.54_{-0.12}^{+0.22}$&$0.43\pm 0.02$&$29.34\pm 0.89$&$23.42\pm 0.09$&$1774/1688$\\
6&$8.63_{-0.14}^{+0.18}$&$0.37\pm 0.02$&$28.89\pm 0.89$&$24.73\pm 0.10$&$1708/1688$\\
7&$8.70\pm 0.18$&$0.36\pm 0.02$&$28.24\pm 0.99$&$24.11\pm 0.09$&$1761/1687$\\
8&$8.99\pm 0.19$&$0.36\pm 0.02$&$29.79_{-0.99}^{+0.96}$&$21.91\pm 0.08$&$1829/1688$\\
9&$9.31\pm 0.20$&$0.36\pm 0.02$&$27.52\pm 1.05$&$18.65\pm 0.07$&$1791/1688$\\
10&$9.57\pm 0.23$&$0.39\pm 0.02$&$30.59\pm 1.11$&$14.65\pm 0.07$&$1815/1687$\\
11&$9.32\pm 0.26$&$0.37\pm 0.03$&$28.80\pm 1.23$&$12.56\pm 0.06$&$1736/1688$\\
12&$9.41\pm 0.29$&$0.32\pm 0.03$&$29.07_{-1.54}^{+1.66}$&$11.73\pm 0.06$&$1810/1688$\\
13&$9.93\pm 0.21$&$0.38\pm 0.03$&$29.18_{-1.51}^{+1.57}$&$11.31\pm 0.07$&$1836/1688$\\
14&$9.40\pm 0.33$&$0.43\pm 0.04$&$29.52\pm 1.45$&$10.94\pm 0.07$&$1863/1688$\\
15&$8.91\pm 0.36$&$0.39\pm 0.04$&$30.49\pm 1.45$&$12.84\pm 0.08$&$1726/1688$\\
16&$9.73\pm 0.39$&$0.37\pm 0.04$&$30.81_{-1.78}^{+1.81}$&$11.78\pm 0.08$&$1742/1688$\\
17&$9.09\pm 0.42$&$0.26\pm 0.04$&$31.58_{-2.80}^{+2.67}$&$10.62\pm 0.07$&$2019/1688$\\
\\ 
 \hline
\end{tabular}
\end{table}
 
   \begin{figure}
   \centering
\includegraphics[width=0.46\textwidth]{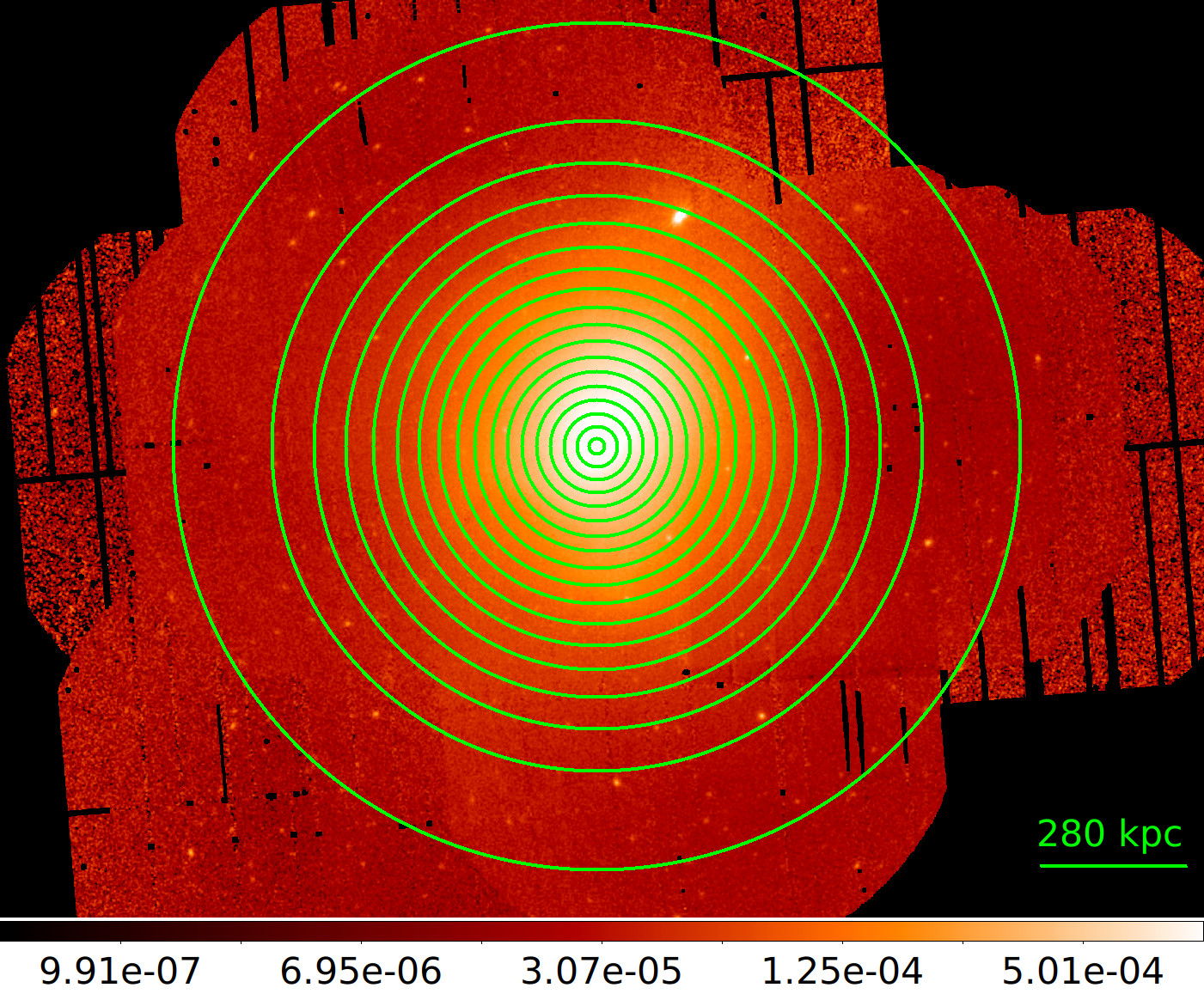}\\
\includegraphics[width=0.46\textwidth]{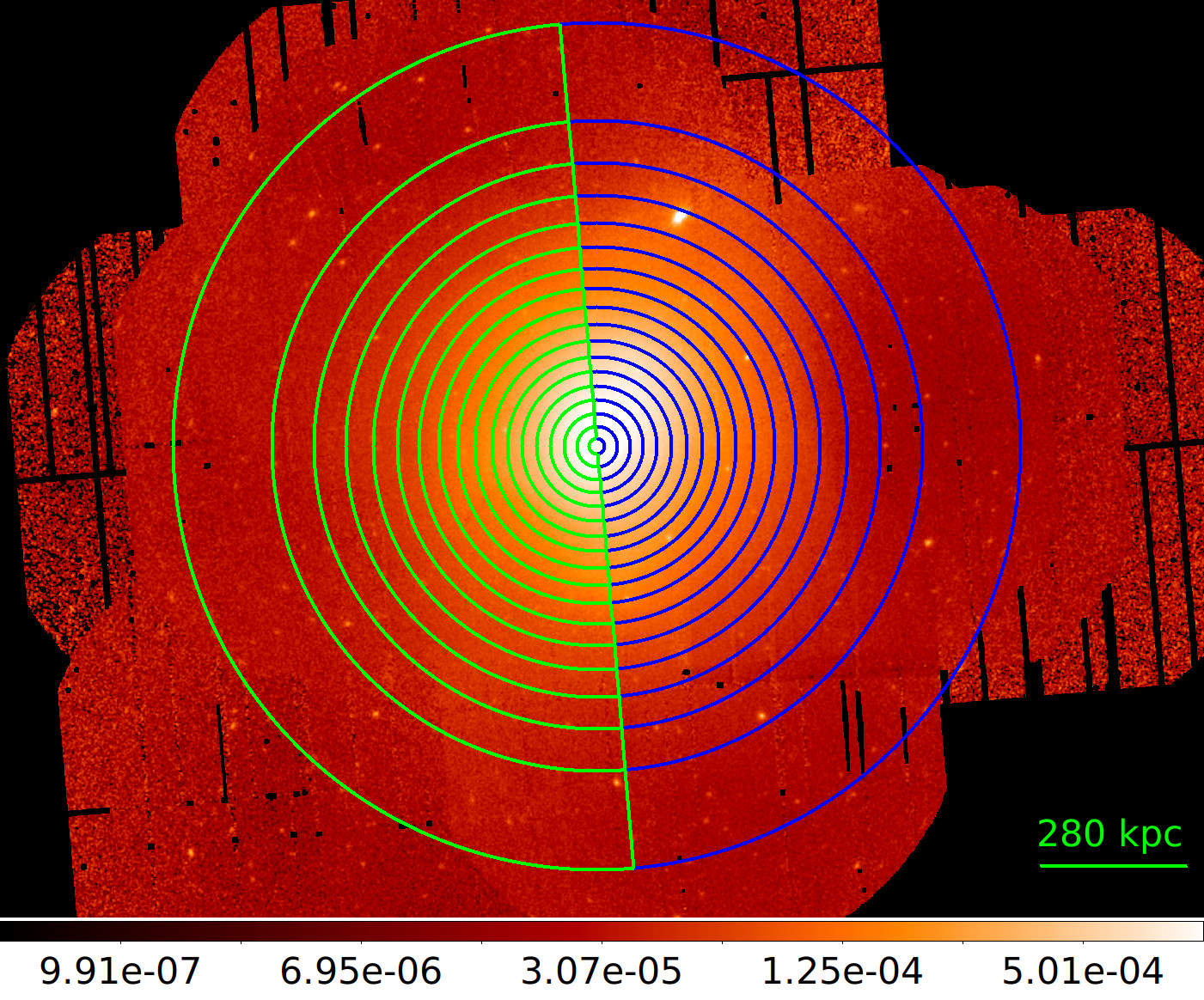}
\caption{Ophiuchus annular regions analyzed for Case 1 (top panel) and Case 2 (bottom panel, see Section~\ref{circle_rings}). } \label{fig_cas1_region} 
    \end{figure}

   \begin{figure*}
   \centering
\includegraphics[width=1.0\textwidth]{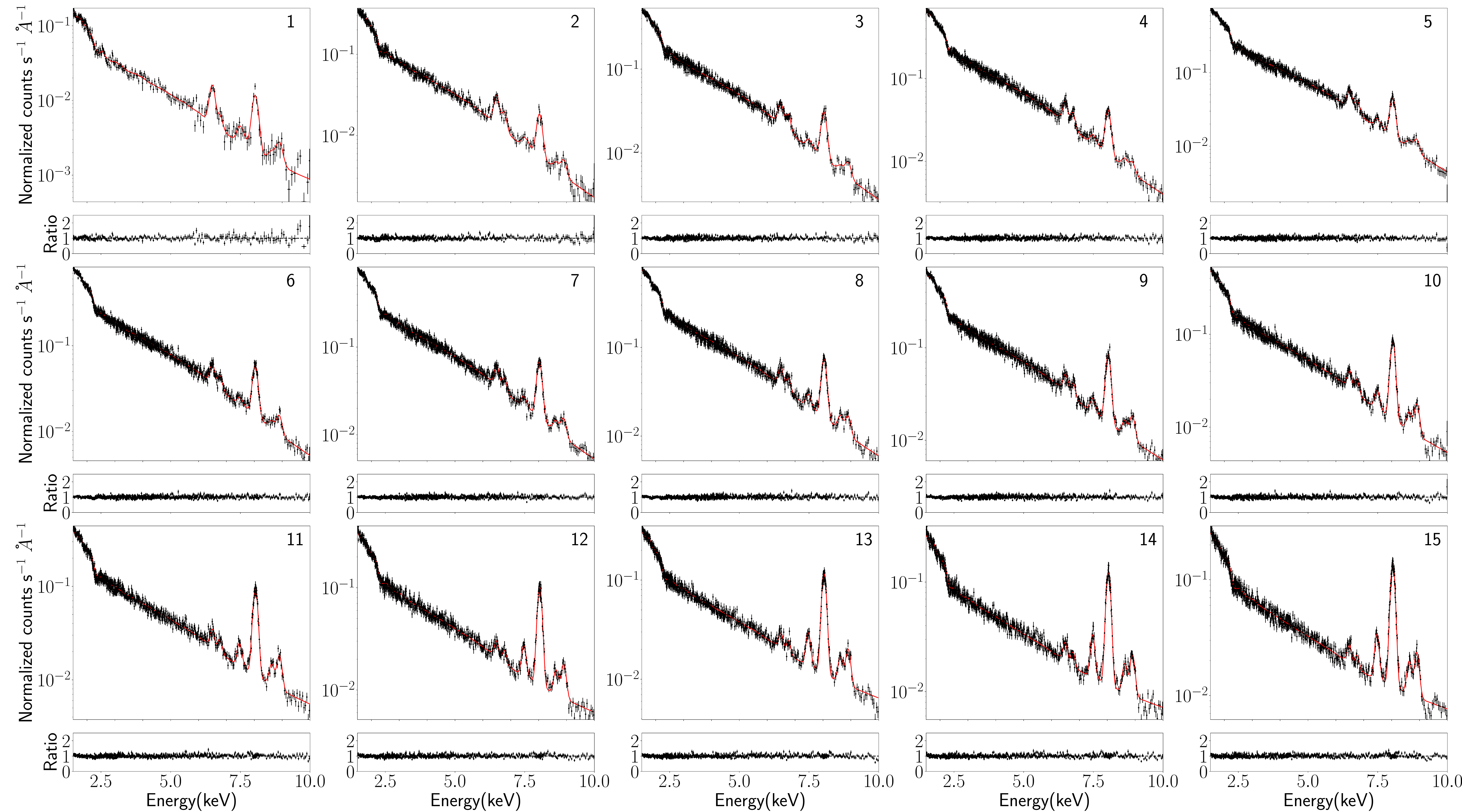}
\caption{A sample of best-fit results obtained for Case 1 (see Section~\ref{circle_rings}). The data have been rebinned for illustrative purposes. } \label{fig_cas1_spectraa} 
    \end{figure*}

\begin{figure} 
\centering 
\includegraphics[width=0.46\textwidth]{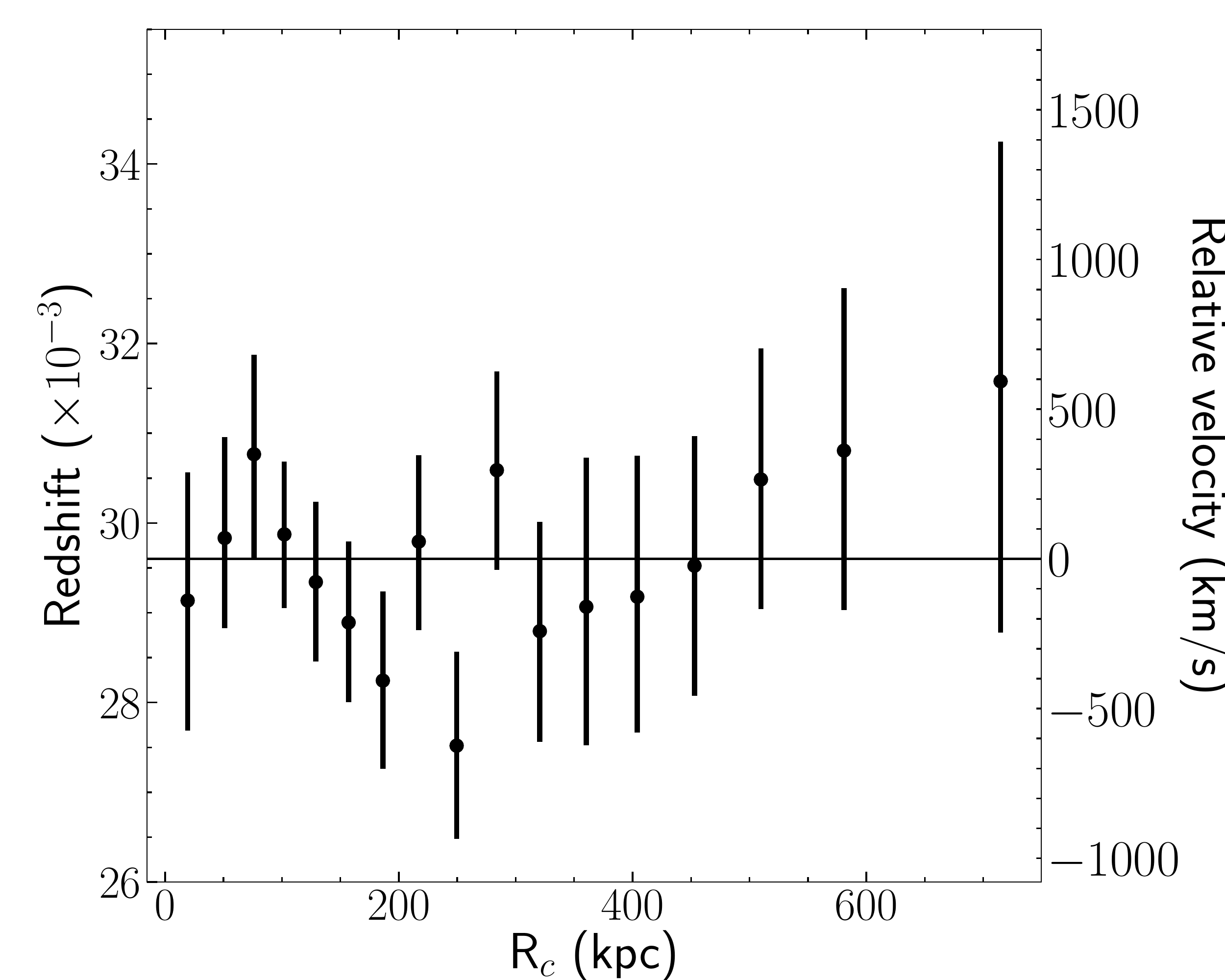} 
        \caption{Velocities obtained for each region for Case 1 (numbered from the center to the outside). The Ophiuchus redshift is indicated with a horizontal line. } \label{fig_cas1_resultsa} 
\end{figure}

\begin{figure} 
\centering  
\includegraphics[width=0.46\textwidth]{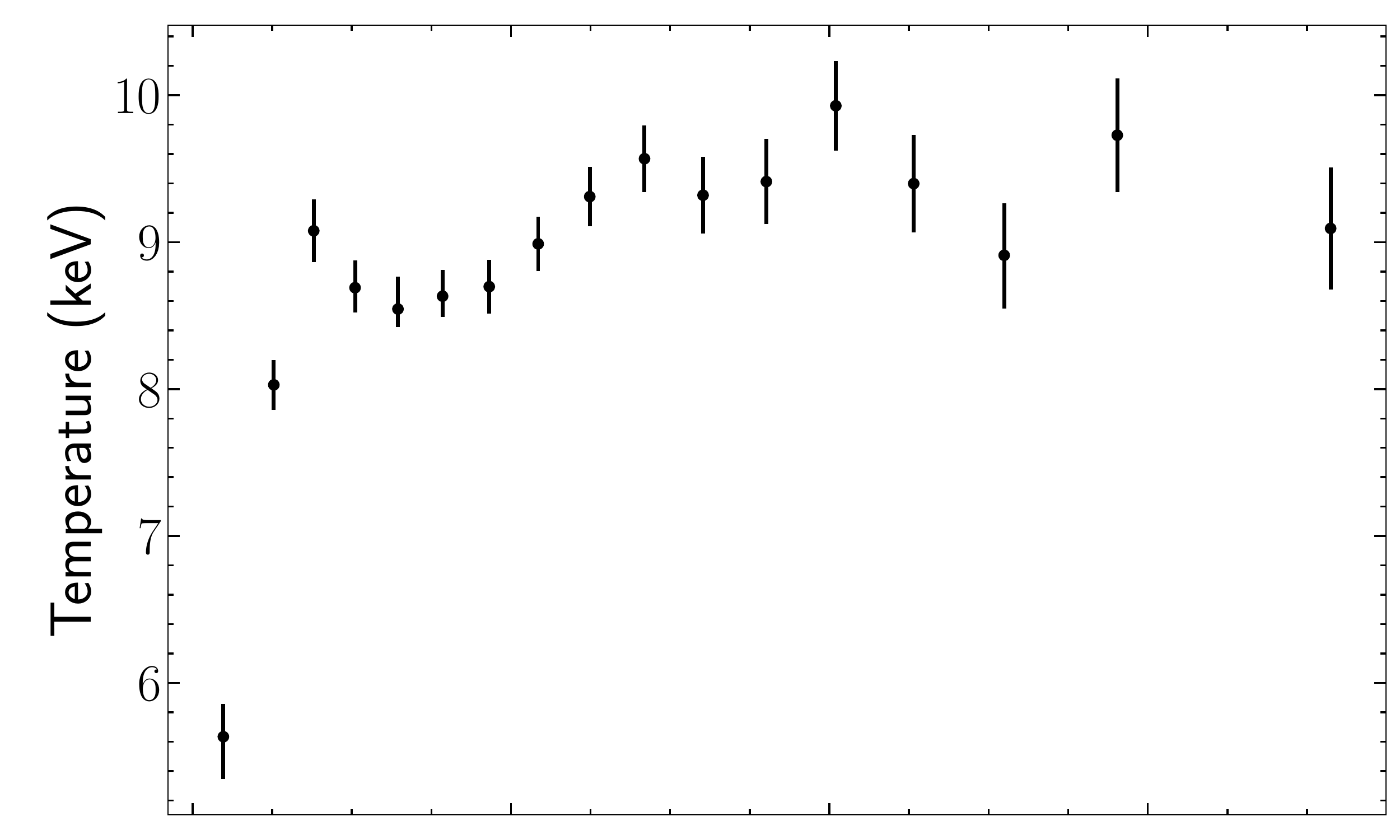}\\
\includegraphics[width=0.46\textwidth]{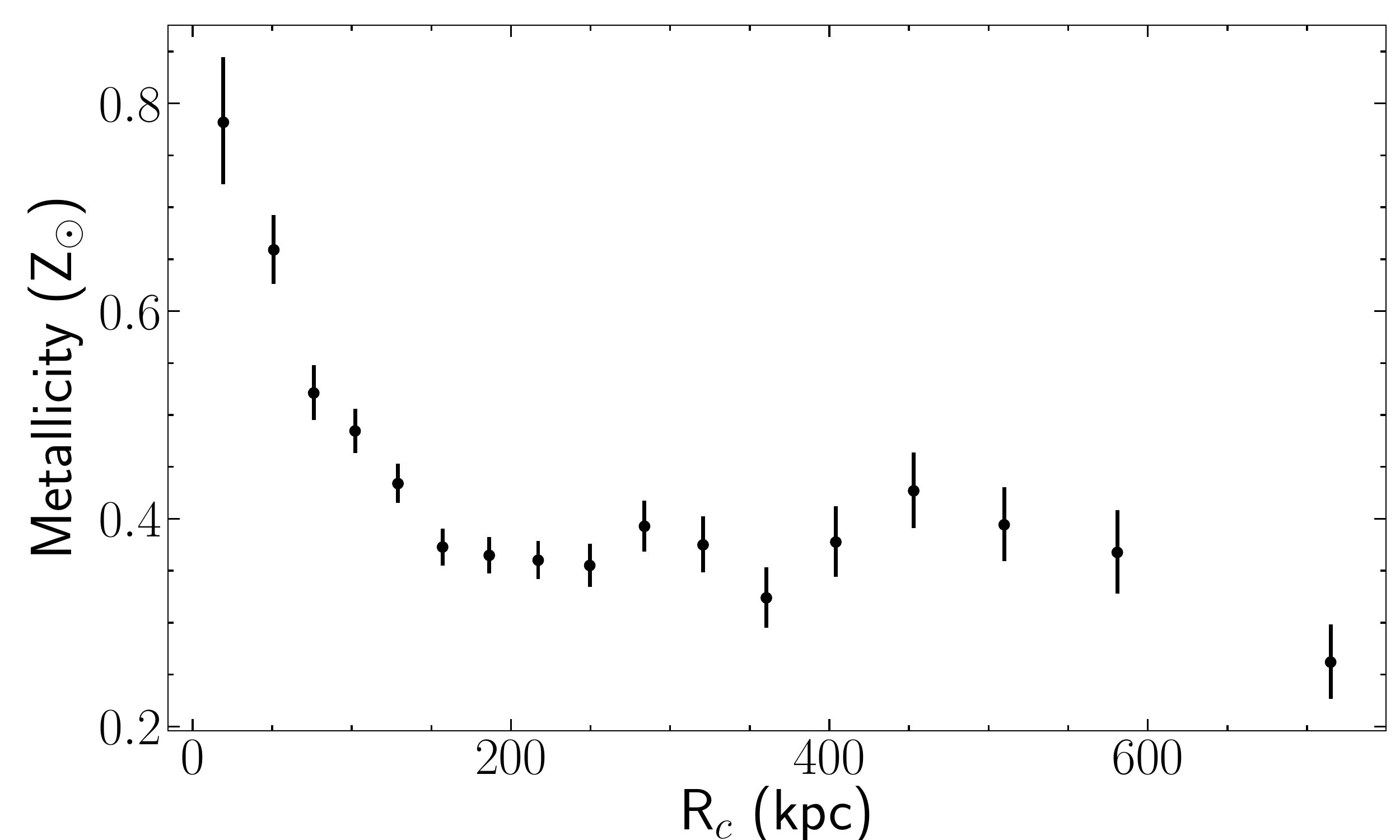} 
        \caption{Temperature (top panel) and metallicity (bottom panel) profiles obtained from the best fit results for Case 1.} \label{fig_cas1_resultsb} 
\end{figure}

\begin{figure} 
\centering 
\includegraphics[width=0.46\textwidth]{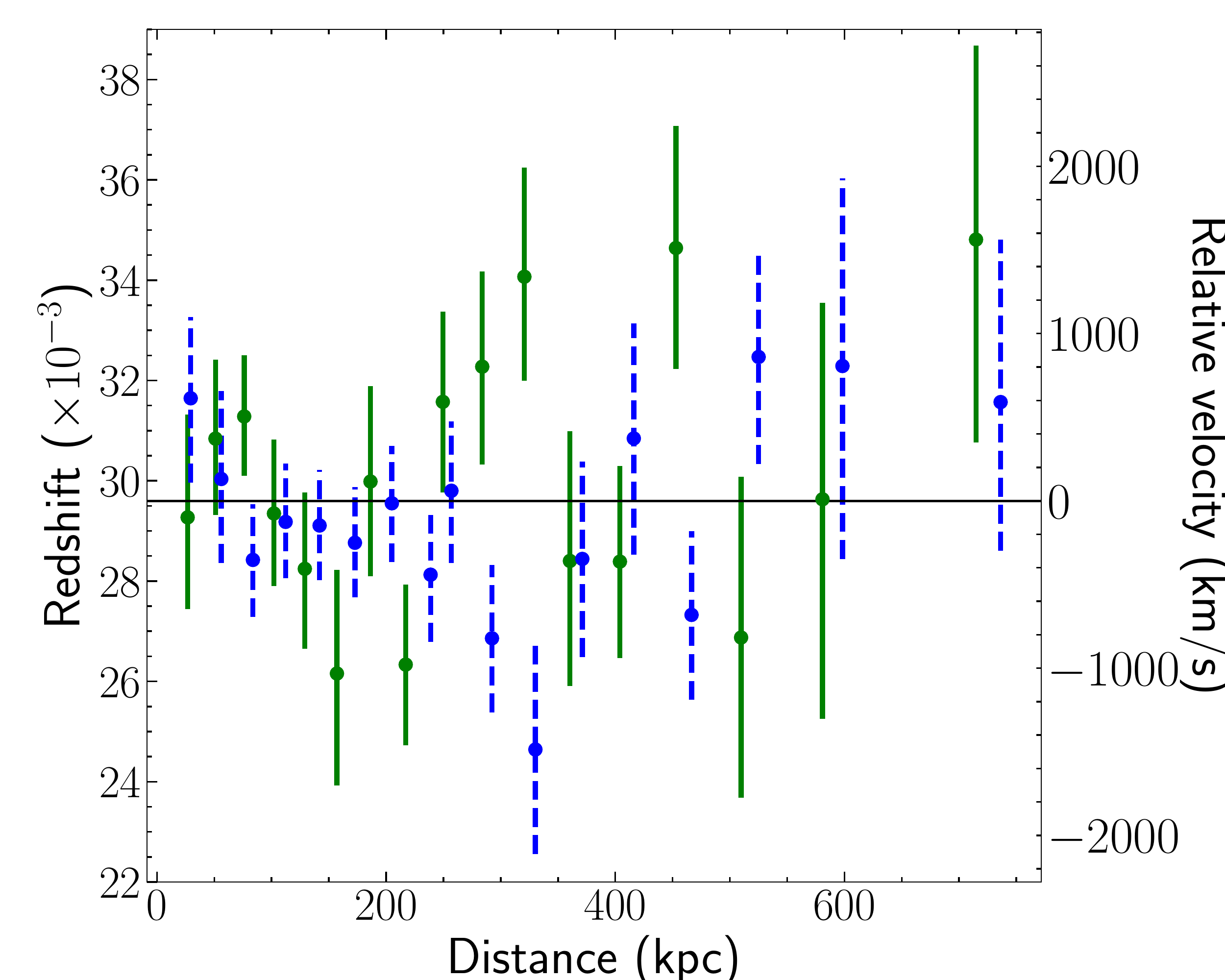} 
        \caption{Velocities obtained for each region for Case 2 (numbered from the center to the outside). The Ophiuchus redshift is indicated with an horizontal line. Green points correspond to results obtained for regions in the E direction while blue dotted correspond to regions in the W direction. Data points correspond to the same distance but have been displaced for illustrative purposes.} \label{fig_cas2_resultsa} 
\end{figure}

\begin{figure} 
\centering  
\includegraphics[width=0.46\textwidth]{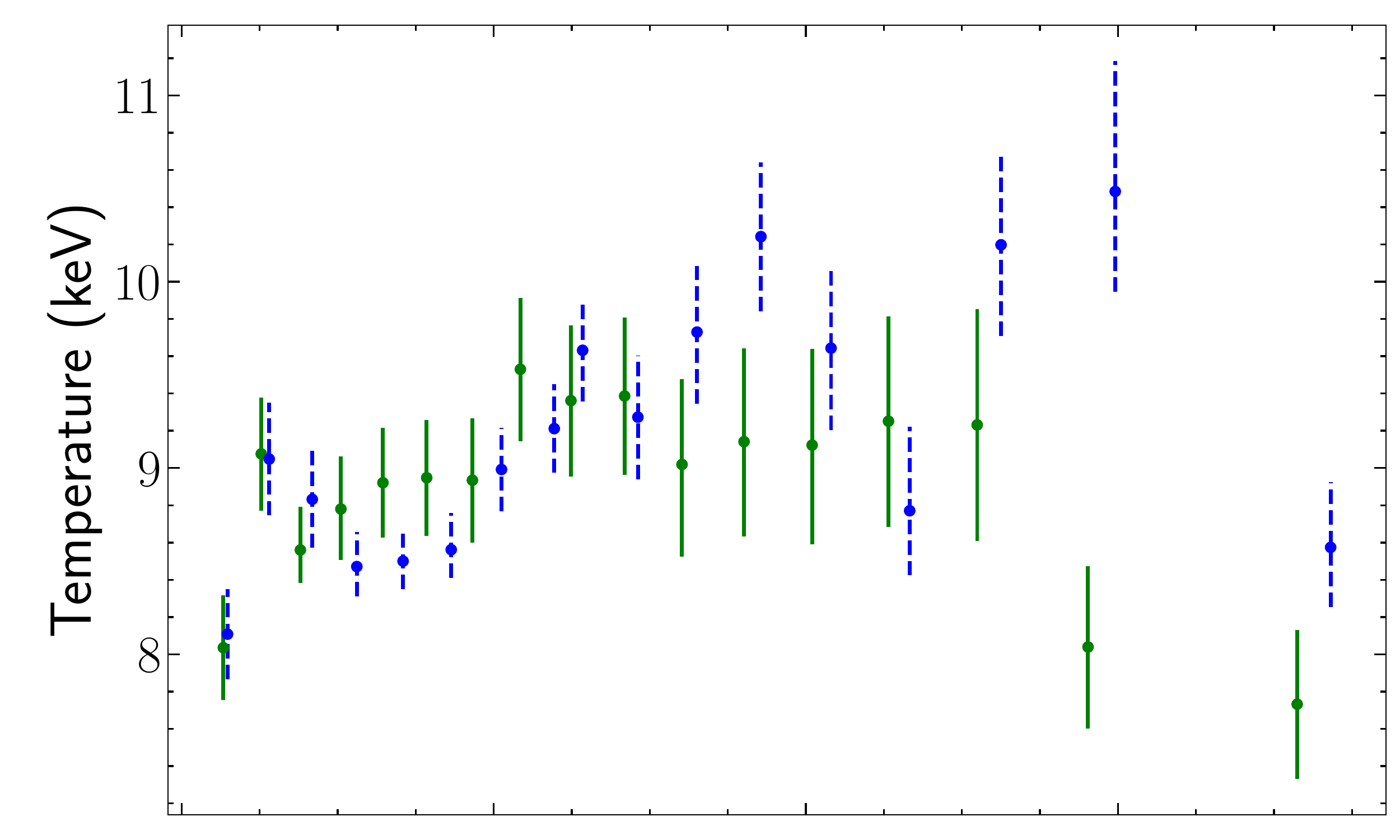}\\ 
\includegraphics[width=0.46\textwidth]{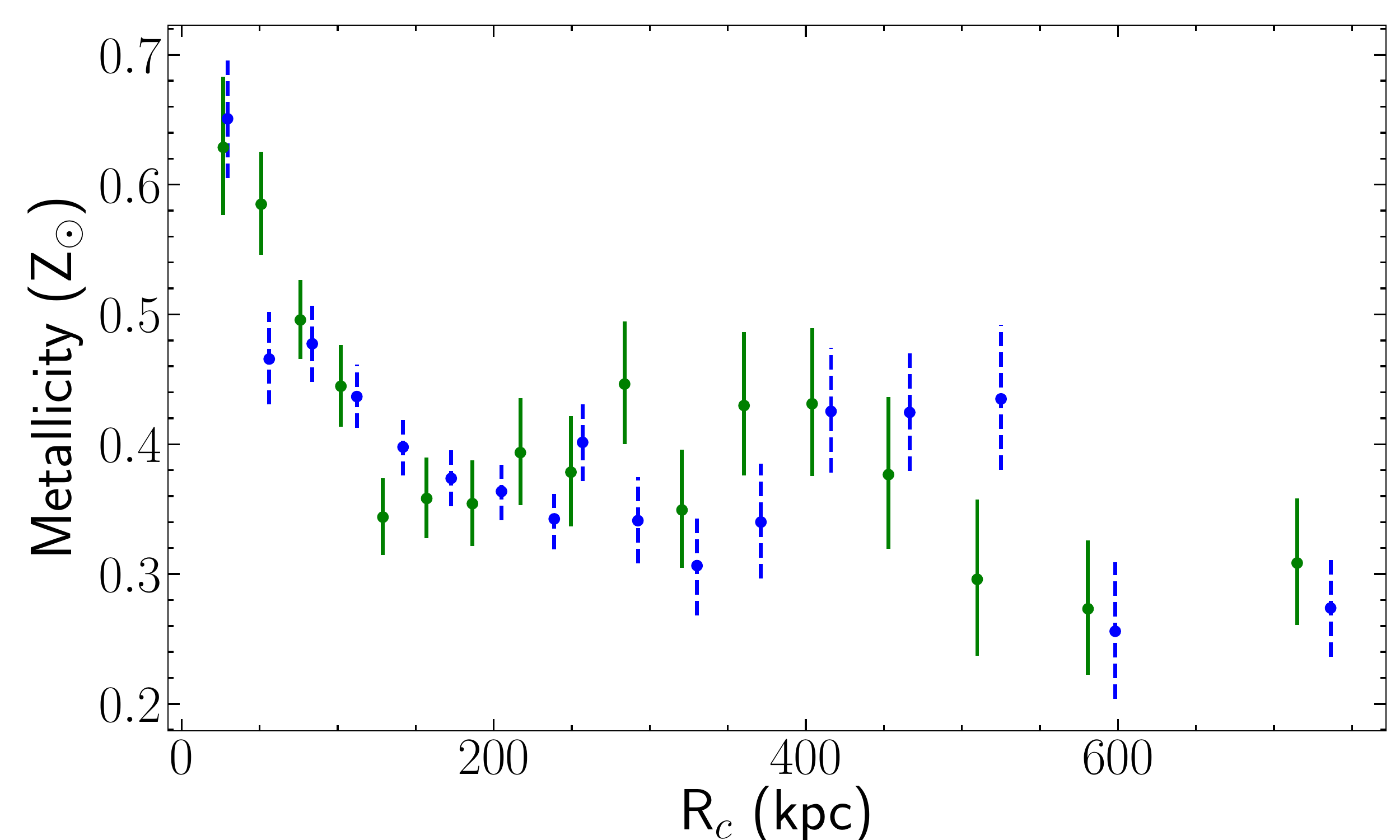} 
        \caption{Temperature (top panel) and metallicity (bottom panel) profiles obtained from the best fit results for Case 2. Green points correspond to results obtained for regions in the E direction while blue dotted correspond to regions in the W direction.} \label{fig_cas2_resultsb} 
\end{figure}

\section{XMM-Newton observations}\label{sec_dat} 

In order to measure the velocity structure within the Ophiuchus cluster by applying the technique described in \citet{san20}, 13 observations were taken between September 2021 and March 2022 (PI: E. Gatuzz, ID: 88028). We also included in our analysis 12 additional observations available in the archive. Table~\ref{tab_obsids} describes the observations, including IDs, coordinates, dates and exposure times. We reduced the {\it XMM-Newton} European Photon Imaging Camera \citep[EPIC,][]{str01} spectra with the Science Analysis System (SAS\footnote{\url{https://www.cosmos.esa.int/web/xmm-newton/sas}}, version 19.1.0). All observations were processed with the {\tt epchain} SAS tool, including only single-pixel events (PATTERN==0) and filtering the data with FLAG==0 to avoid regions close to CCD edges or bad pixels. Bad time intervals were filtered from flares applying a 1.0 cts/s rate threshold. 

After obtained the event files we follow the technique described in \citet{san20} in order to calibrate the absolute energy scale of the EPIC-pn detector using the instrumental X-ray emission lines identified in the spectra of the detector. In this way corrections are applied to (1) the average gain of the detector during the observation, (2) the spatial gain variation across the detector with time and (3) the energy scale as a function of detector position and time. In particular, we used the updated calibration files computed by \citet{gat22a}. For each observation, we created the final event files after applying this energy scale correction.

The X-ray image obtained in the 0.5-9.25~keV energy band is shown in Figure~\ref{fig_xray_maps} (left panel). We used the SAS task {\tt edetect\_chain} with a likelihood parameter {\tt det\_ml} $> 10$ to identify point-like sources from the image, which were excluded in the following analysis. The exposure map in the 0.5-9.25~keV energy range is shown in Figure~\ref{fig_xray_maps} (top right panel). Figure~\ref{fig_xray_maps} bottom panel shows the number of counts in each 1.59 arcsec pixel in the Fe-K complex (6.50 to 6.90 keV, in restframe), after subtracting neighboring scaled continuum images (i.e. total number of counts in 6.06 to 6.43 and 6.94 to 7.16 keV energy ranges, in restframe). A gaussian smoothing of $\sigma = 3$ pixels was applied for illustrative purposes. The image shows that a large number of photons are obtained around the cluster center due to the spatial offset in the observations which allows us to cover the Cu-K$\alpha$ hole of the detector.

 Figure~\ref{fig_xra_sb} shows the fractional difference in 0.5 to 9.25 keV surface brightness from the average at each radius. The black regions correspond to bad pixels from the detector whose sizes have been artificially increased due to the image smoothing. There are two clear sharp discontinuities along the NW and SE directions at $\sim 260-280$~kpc from the cluster center who are detected in this work for the first time. In order to check whether such edges are due to variations in exposure across the field of view  (e.g. the location of chip edges observed in Figure~\ref{fig_xray_maps} top right panel) we created a surface brightness map using the observation ObsID=0505150101 (see Table~\ref{tab_obsids}) which points to the cluster center and covers an area larger than the discontinuities. We have obtained the same structure in such image, pointing out that such surface brightness edges are real. We analyzed this structure in detail in subsection~\ref{regions_cold_fronts}.

\section{Spectral fitting procedure}\label{sec_fits} 
For each spatial region analyzed, we combine the spectra from different observations together and rebinned the spectra to have at least 1 count per channel. We load the data twice in order to fit separately, but simultaneously, the hard 4.0-10 keV and the softer 1.5-4.0 keV energy bands. As described in \citet{gat22b}, the inclusion of the lower energy band leads to a much better constraint for the temperatures and metallicities. However, the higher temperature of the plasma in the Ophiuchus cluster, as compared to Virgo and Perseus, makes it slightly harder to measure velocities owing to smaller equivalent widths of the Fe K lines. We model the cluster gas emission with a simple {\tt apec} model. Previous studies of the Ophiuchus cluster have shown the presence of a multi-temperature component near the cluster center \citep{nev09,mil10,wer16b}. However, because we are not including the soft energy band in our analysis we decide to model one component. Moreover, when using the multi-temperature {\tt lognorm} model \citep{gat22b} we found similar velocities to those obtained with the {\tt apec} model, besides the spectral resolution below $\sim$1~keV is significantly lower in EPIC-pn (see Appendix~\ref{sec_single_apec}).  In order to account for the Galactic absorption we included a {\tt tbabs} component \citep{wil00}.

The free parameters in the model are the temperature, metallicity (i.e. total metal abundances) and normalization. Given that for lower energies the new EPIC-pn energy calibration scale cannot be applied, the redshift is a free parameter only for the 4.0-10 keV energy band as in \citet{gat22b}. We fixed the column density to $3.40\times 10^{21}$ cm$^{-2}$ \citep{wil13}, although it is important to note that in the energy range analyzed the absorbing component has a weak effect in the velocity measurement, which depends on the Fe K$\alpha$ energy. It is very likely that there is a non-uniform distribution of the absorbing column densities due to the large angular scale of the observations and the proximity to the Galactic plane. A detailed study of the $N({\rm H})$ impact in the temperature-metallicity measurements will be carried out in a later work which will include the $<1.5$~keV energy range.

We analyze the spectra with the {\it xspec} spectral fitting package (version 12.11.0k\footnote{\url{https://heasarc.gsfc.nasa.gov/xanadu/xspec/}}). We assumed {\tt cash} statistics \citep{cas79}. Errors are quoted at 1$\sigma$ confidence level unless otherwise stated. \citet{gat22a} discussed about systematic uncertainties in the measurements when applying this method (Section~4.7). As background components we have included Cu-$K\alpha$, Cu-$K\beta$, Ni-$K\alpha$ and Zn-$K\alpha$ instrumental emission lines, and an instrumental powerlaw component with its photon index fixed at 0.136, a value derived from the analysis of all observations in the archive done by \citep{san20}. Finally, abundances are given relative to \citet{lod09}.

\subsection{Velocity radial profiles}\label{circle_rings}
We analyzed the velocity radial profile by fitting X-ray spectra from annular regions for two cases:
\begin{itemize}
\item \emph{Case 1:} complete concentric rings, square root spaced and with the center located at the cluster center. 
\item \emph{Case 2:} concentric regions divided in E-W zones, square root spaced and with the center located at the cluster center.  
\end{itemize}

The regions analyzed in Case 1 are shown in Figure~\ref{fig_cas1_region} (top panel) while the best fit results are listed in Table~\ref{tab_circular_fits}. Figure~\ref{fig_cas1_spectraa} shows the best-fit spectra for the first 15 regions. Figures~\ref{fig_cas1_resultsa} and~\ref{fig_cas1_resultsb} show the velocities, temperatures and metallicities obtained from the best-fit per region.  We have obtained velocities measurements with uncertainties down to $\Delta v\sim 240$ km/s (for ring 4). The error-weighted standard deviation of the sample is $\sigma=319$~km/s.  While most of velocities show no deviation from the cluster velocity, we have found a significant change in velocities from blueshifted to redshifted gas (with respect to the Ophiuchus cluster) located at $\sim 250$ kpc, going from $-623\pm 312$~km/s to $296\pm 330$~km/s. This corresponds to a departure from systemic at about the $2\sigma$ confidence level. The Ophiuchus cluster is know by the high temperature of the gas within it \citep{nev09,wer16}.   As indicated in Appendix~\ref{sec_single_apec}, we tested a multi-temperature component. With such a model we found better fits, from the statistics point of view. However, it is important to note that our spectra analysis does not include the soft energy band, hence the temperature could be over(under)estimated. A detailed study of the metallicity and temperature distribution is beyond the scope of this analysis, which is focused on the velocity structure of the ICM.

The regions analyzed for Case 2 are show in Figure~\ref{fig_cas1_region} (bottom panel), while Figures~\ref{fig_cas2_resultsa} and~\ref{fig_cas2_resultsb} show the temperatures, metallicities and velocities obtained from the best-fit per region. The regions were selected given the symmetry on the temperature and metallicities found along this directions by \citet{wer16}. The error-weighted standard deviation of the samples are $\sigma=843$~km/s (towards E) and $\sigma=661$~km/s (towards W). In these plots blue dotted symbols correspond to regions towards W direction while green points correspond to regions towards the E direction. In some cases, the same region along different directions displays opposite velocities (e.g. regions 3,10,11,14,15 from left to right). The velocities for some adjacent regions suddenly change from blueshifted to redshifted values, for example from regions 8 to 9 ($\sim$1.4$\sigma$ confidence level),  11 to 12 ($\sim$1.2$\sigma$ confidence level), 13 to 14 ($\sim$1.4$\sigma$ confidence level) and 14 to 15 ($\sim$1.8$\sigma$ confidence level) along the E direction and from regions 14 to 15 ($\sim$1.9$\sigma$ confidence level) along the W direction. We tried fitting the obtained velocities with a flat model and we obtained a $red-\chi^{2}>10$. For distances $<250$~kpc the temperatures along the E direction are larger than those obtained along the W direction while the metallicities are similar while for distances $>250$~kpc the temperatures along the E direction are lower than those along the W direction.

\begin{figure*} 
\centering 
\includegraphics[width=0.47\textwidth]{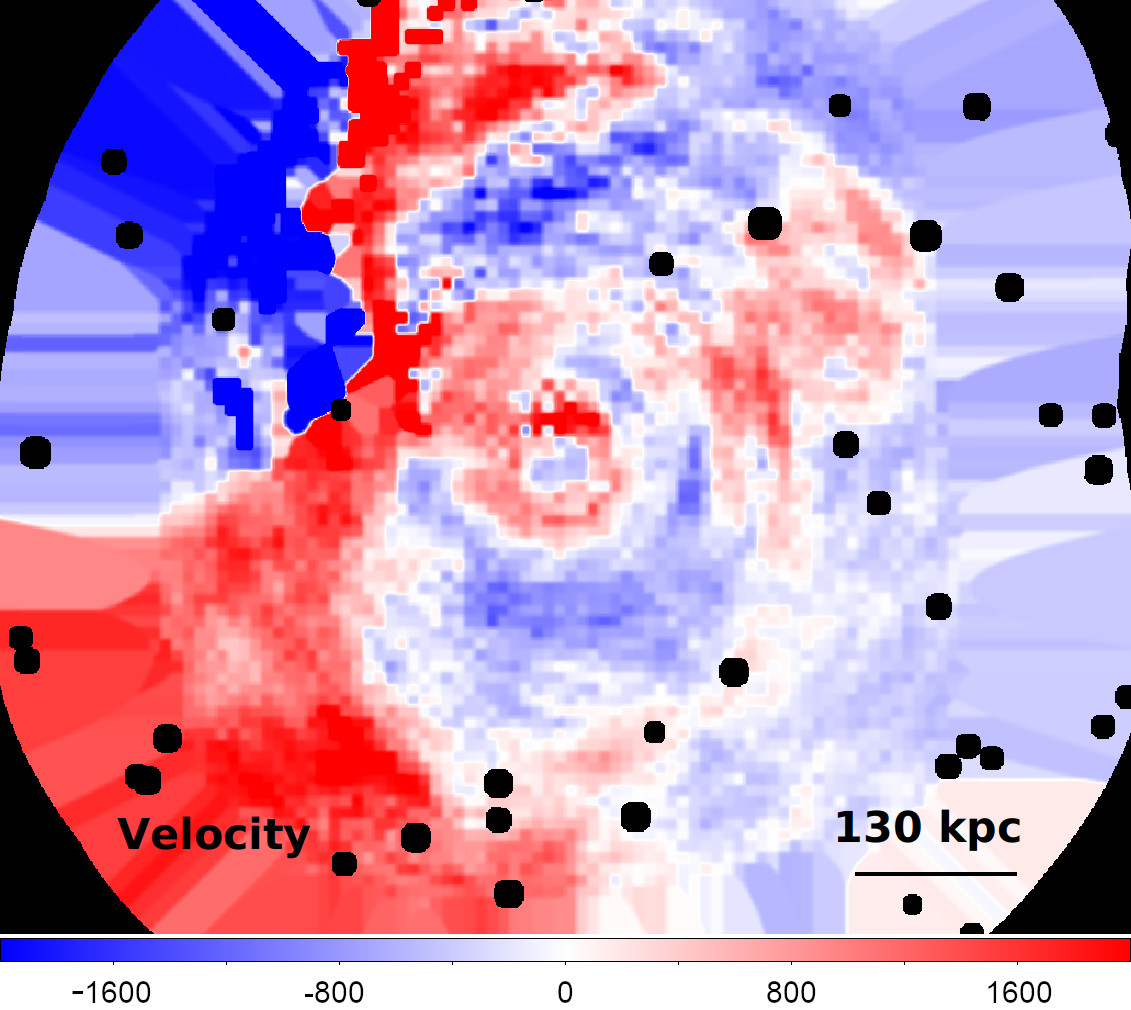} 
\includegraphics[width=0.47\textwidth]{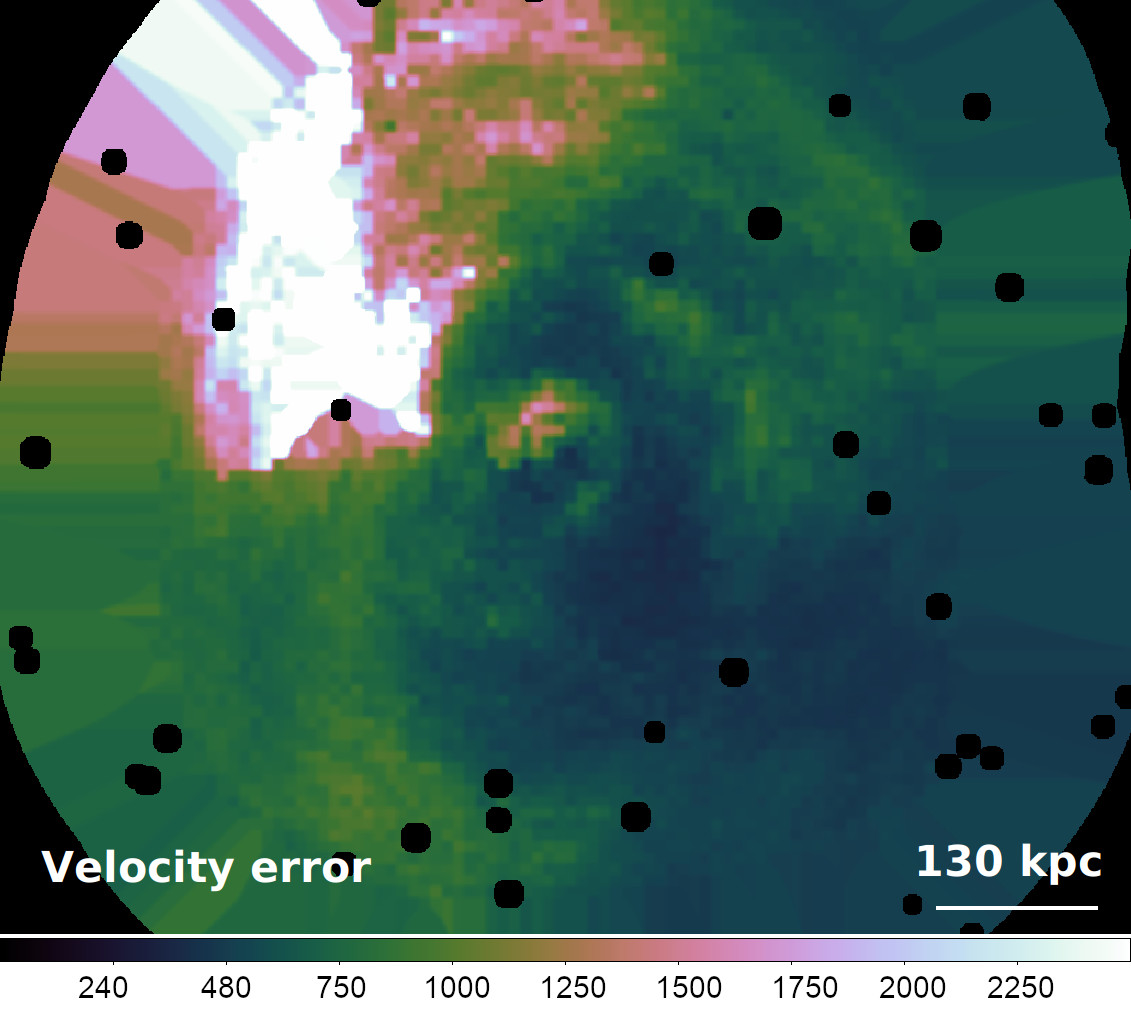}\\ 
\includegraphics[width=0.47\textwidth]{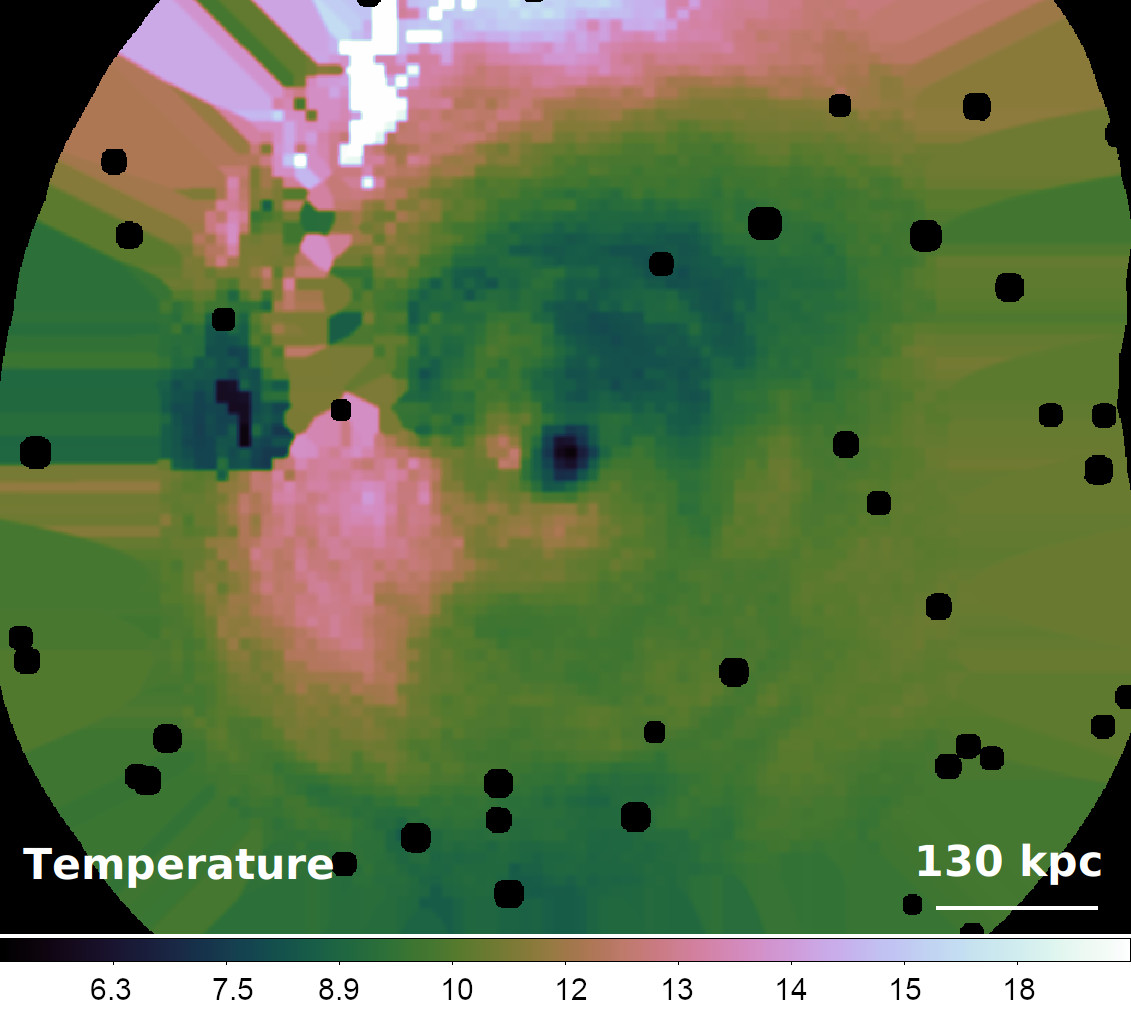}
\includegraphics[width=0.47\textwidth]{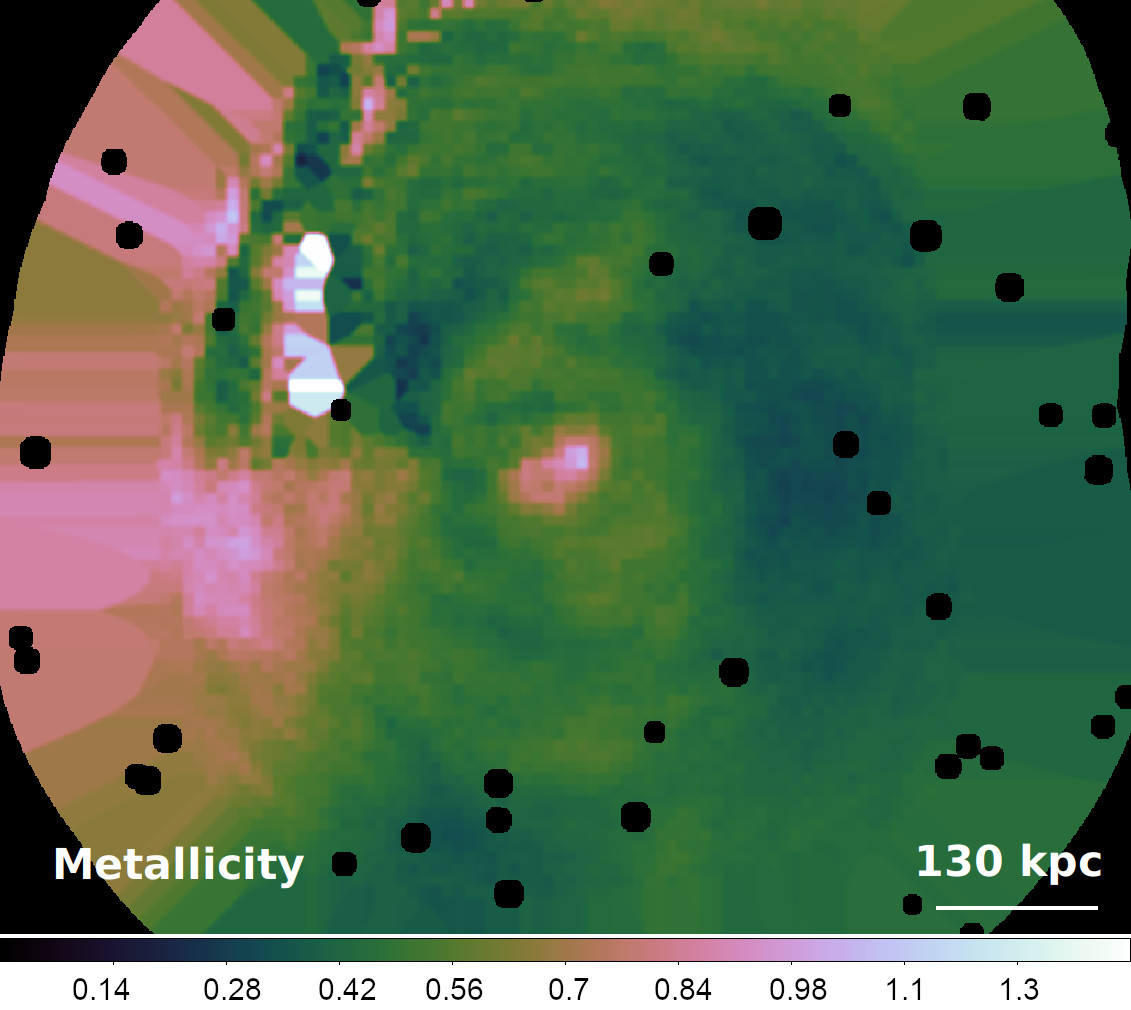}
\caption{\emph{Top left panel:} velocity map (km/s) relative to the Ophiuchus cluster ($z=0.0296$). Maps were created by moving 2:1 elliptical regions (i.e. rotated to lie tangentially to the nucleus) containing $\sim$750 counts in the Fe-K region.  \emph{Top right panel:}  1$\sigma$ statistical uncertainty on the velocity map (km/s). Black circles correspond to point sources which were excluded from the  analysis. Both maps have been smoothed with a $2\sigma$ Gaussian. \emph{Bottom left panel:} temperature map in units of K.  \emph{Bottom right panel:} abundance map relative to solar abundances from \citet{lod09}.} \label{fig_velocity_ellipses1} 
\end{figure*}

 \begin{figure} 
\centering  
\includegraphics[width=0.41\textwidth]{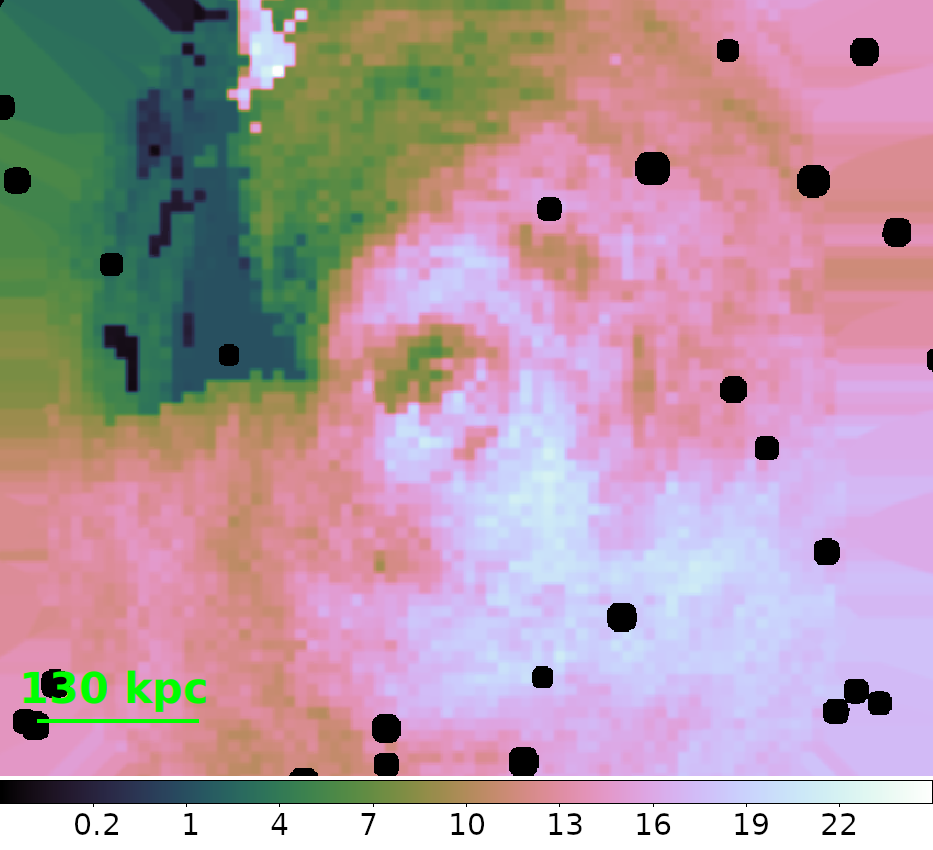}\\
\includegraphics[width=0.41\textwidth]{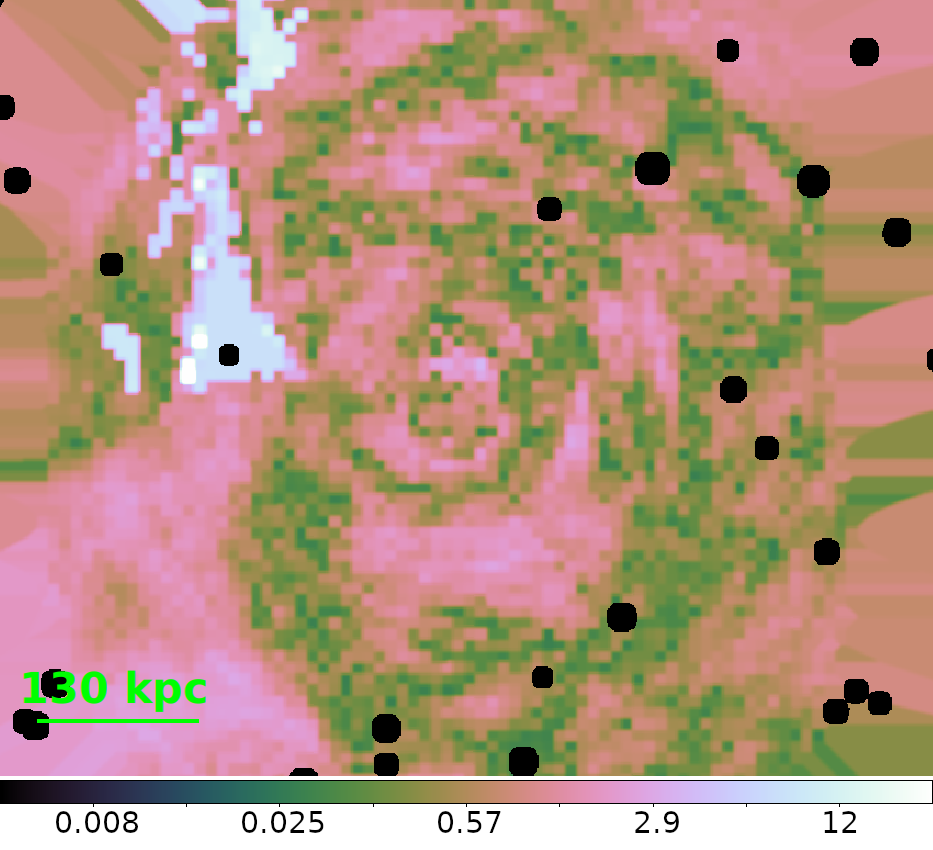}
\caption{\emph{Top panel:} Redshift to redshift error ratio map in linear-scale ($|z/\Delta z|$). \emph{Bottom panel:} Significance map of the measured spectral map redshift in log-scale ($|(z-z_{0})/\Delta z|$, where $z_{0}$ is the Ophiuchus redshift). } \label{fig_vel_velerror_ratio} 
\end{figure}

 \begin{figure} 
\centering  
\includegraphics[width=0.41\textwidth]{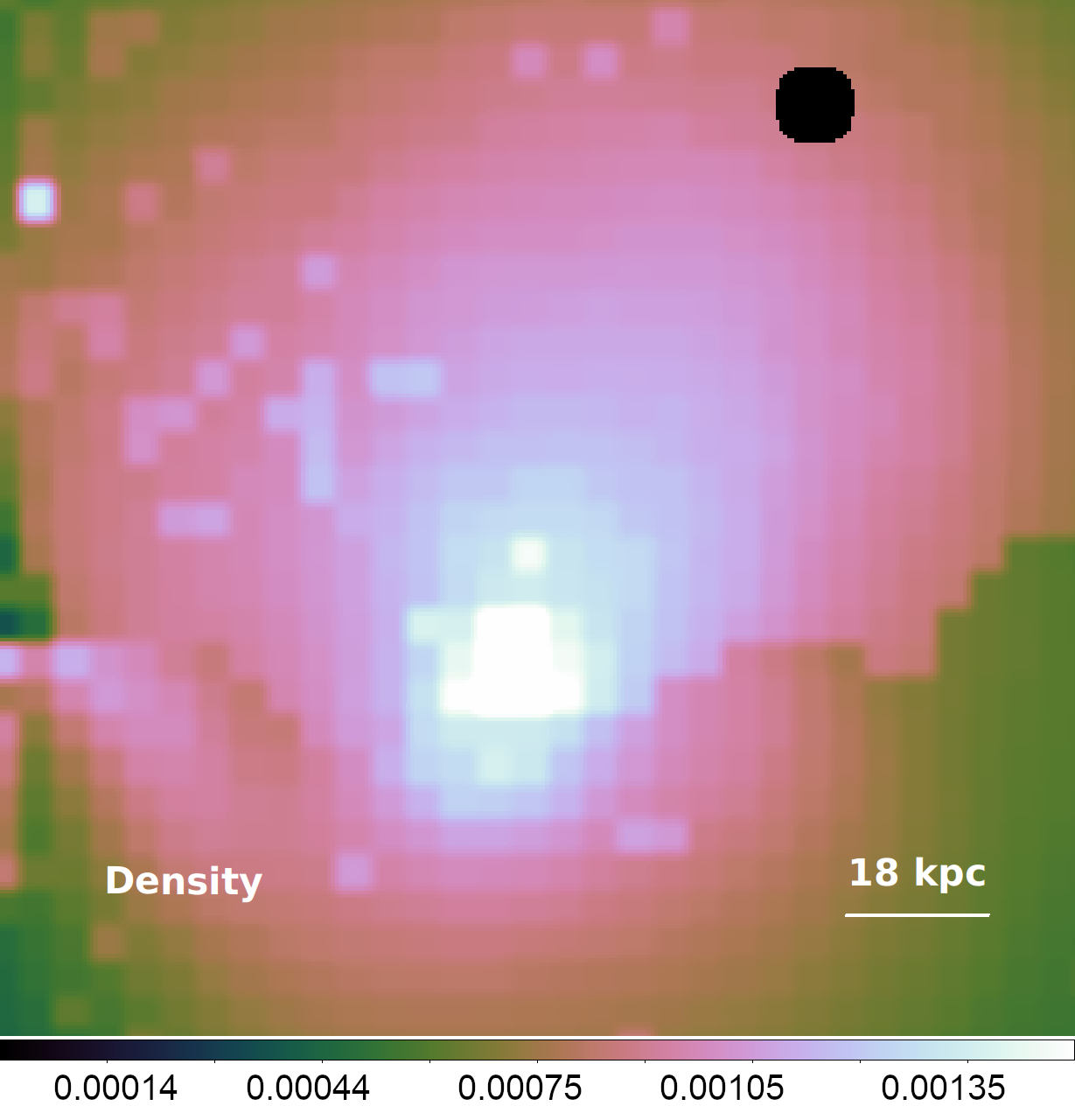}\\
\includegraphics[width=0.41\textwidth]{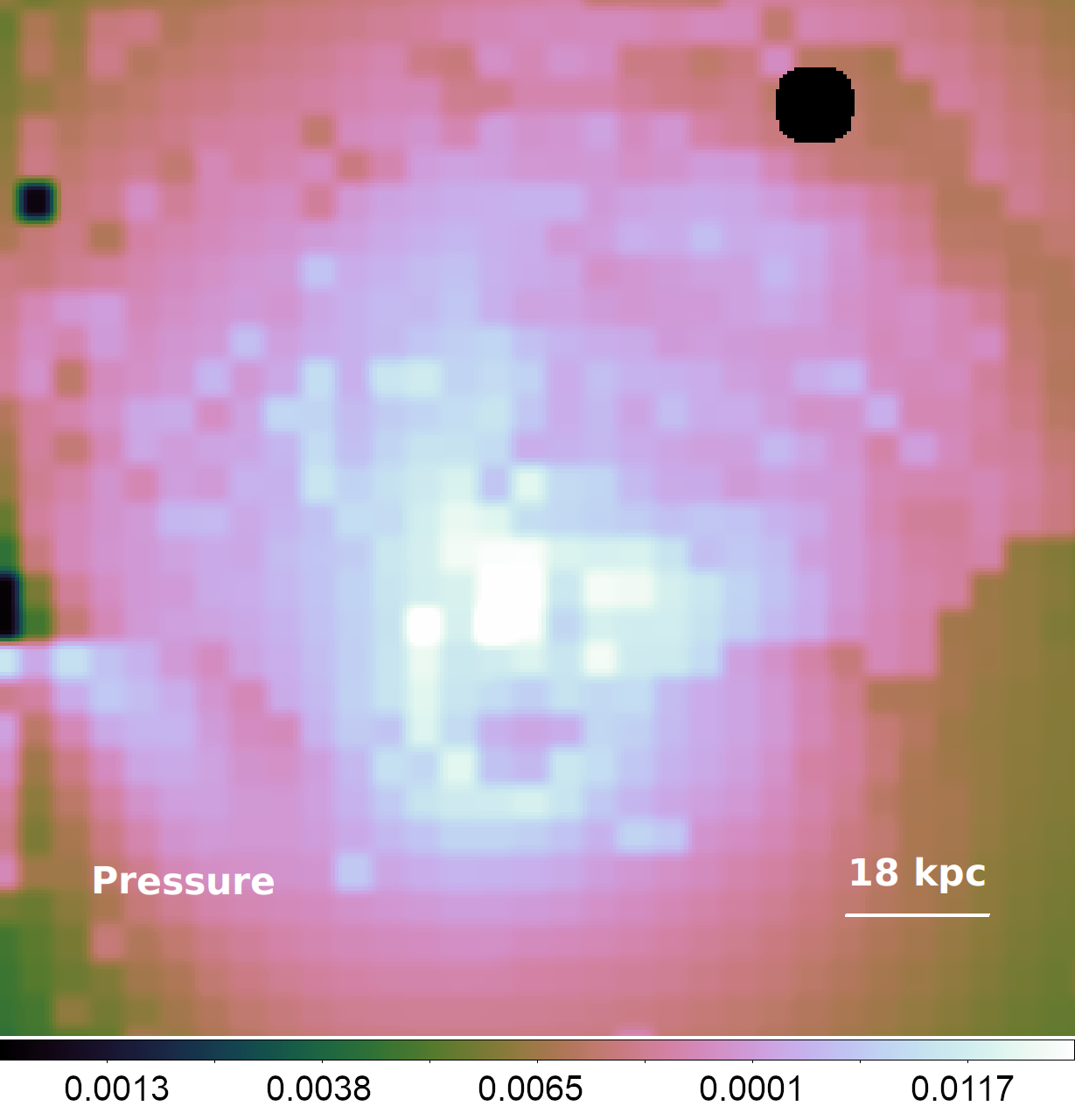}\\
\includegraphics[width=0.41\textwidth]{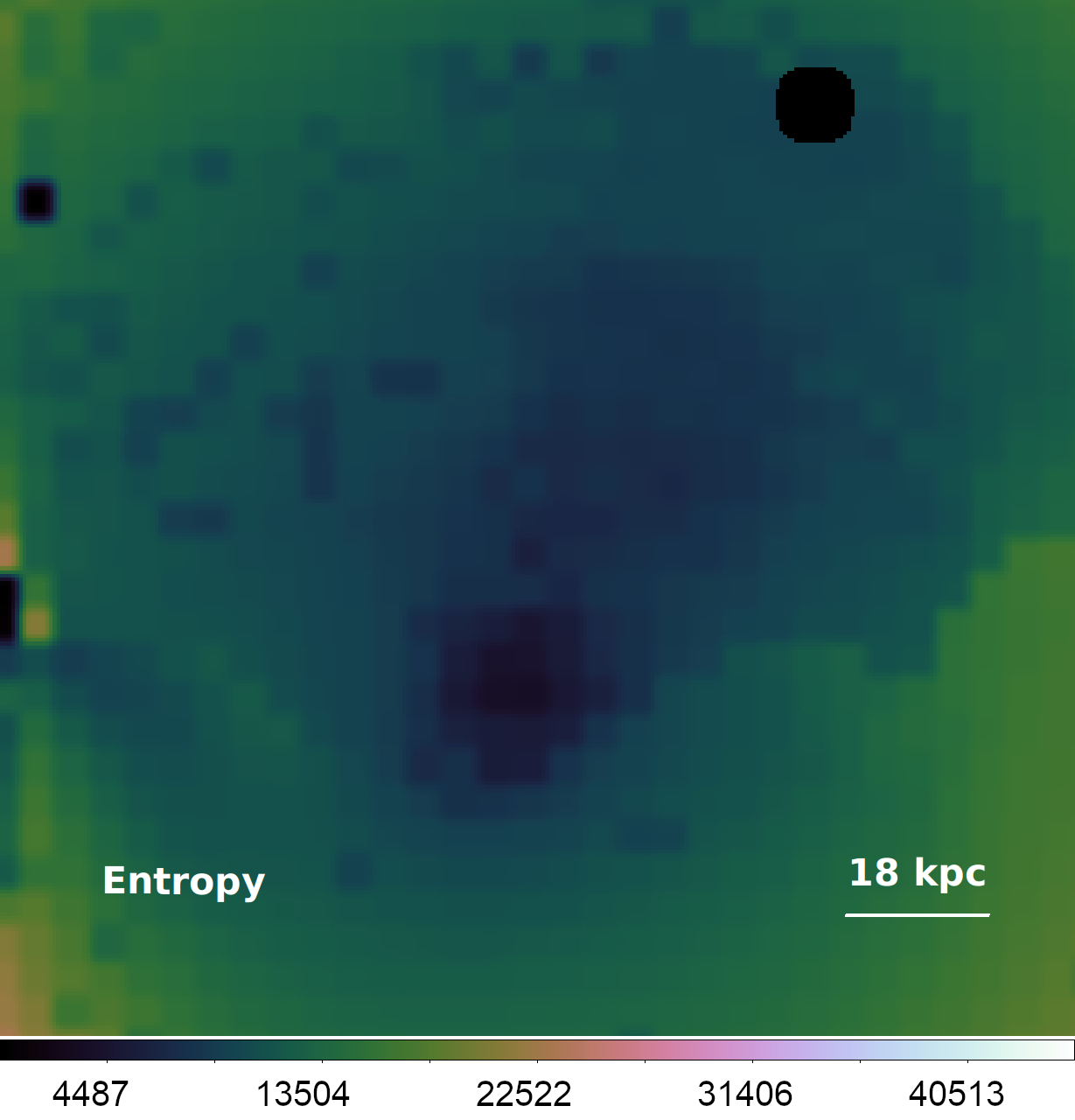}
\caption{\emph{Top panel:} density map (cm$^{-3}$). Black circles correspond to point sources which were excluded from the  analysis. \emph{Middle panel:} pressure map (keV cm$^{-3}$). \emph{Bottom panel:} entropy map (keV cm$^{2}$). } \label{fig_entropy_ellipses} 
\end{figure}

\subsection{Spectral maps}\label{spec_maps} 
Following the method shown in \citet{san20,san22,gat22a,gat22b} we created a velocity map of the cluster. First, we created elliptical regions with a 2:1 axis ratio, rotating them such that the longest axis lay tangentially to a vector pointing to the central core. The radii of the ellipses are adaptively adjusted to make sure each of them have $\sim$750 counts in the Fe-K complex after continuum substraction in order to reduce uncertainties on the velocities. We moved in a grid with a spacing of 0.25 arcmin.  Then, we extracted the spectra for all observations and performed a combined fit for each region.  Finally, we used the same response matrix for all the pixels, but created weighted-average ancillary responses files (arf) for each ellipse, created from the individual ones for each observation.

The velocity map obtained, as well as the uncertainties, are shown in Figure~\ref{fig_velocity_ellipses1} (top panels). In order to improve the clarity we use a red-blue color scale to indicate whether the gas is moving towards (blue) or away from (red) the observer. We did not identified a clear opposite blueshifted-redshifted structure around the cluster center as those identified in the Virgo cluster \citep{gat22b}. Instead, there is a blueshifted region in the cluster core surrounded by a circular-ring of redshifted gas, both regions displaying low velocities. This could be a direct gas sloshing pattern, as those shown in hydrodynamical simulations \citep[e.g., see Figure~19 top middle panel in][]{gat22a}. There is a clear high velocity region along the north-east direction from the cluster center with velocities larger than $2000$ km/s, although the velocity uncertainties are very large in this region. Interestingly, the change between the blueshifted and redshifted gas is striking, showing a discontinuous jump instead of a gradual transition between the two regions. Such structure is not observed in the east direction from the cluster center, where the velocities are better defined in terms of uncertainties. We also found a redshifted region located to the south direction from the cluster core which may be associated with a surface brightness excess identified by \citet{wer16b}.

Figure~\ref{fig_vel_velerror_ratio} shows a significance map of the velocities (i.e. the ratio between the measured redshift and the absolute value of its error). It is clear from the plot that the large uncertainties in velocities (see white region in Figure~\ref{fig_velocity_ellipses1} top right panel) correspond to regions where the significant detection is low. Even when the number of counts is large, the spectra obtained for such regions display a very weak iron complex. Therefore, the best-fit tend to have unrealistic low abundances ($<0.1$), large temperatures ($>50$~keV) and redshift values pegged to the upper/lower limits. Future work, including the soft energy band, may help to determine whether such low iron line presence is due to temperature or abundance effects.

\begin{figure} 
\centering 
\includegraphics[width=0.42\textwidth]{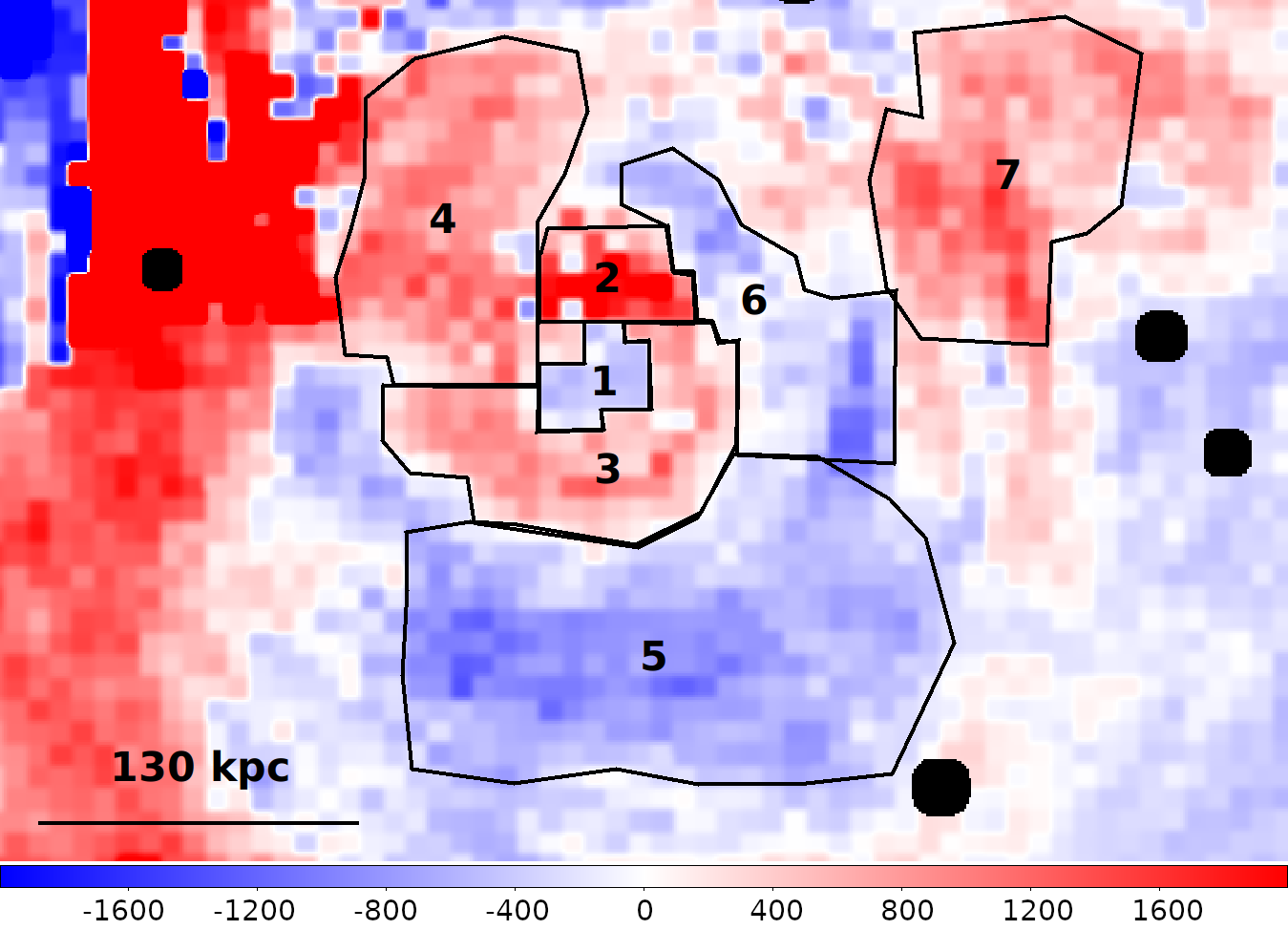}\\
\includegraphics[width=0.42\textwidth]{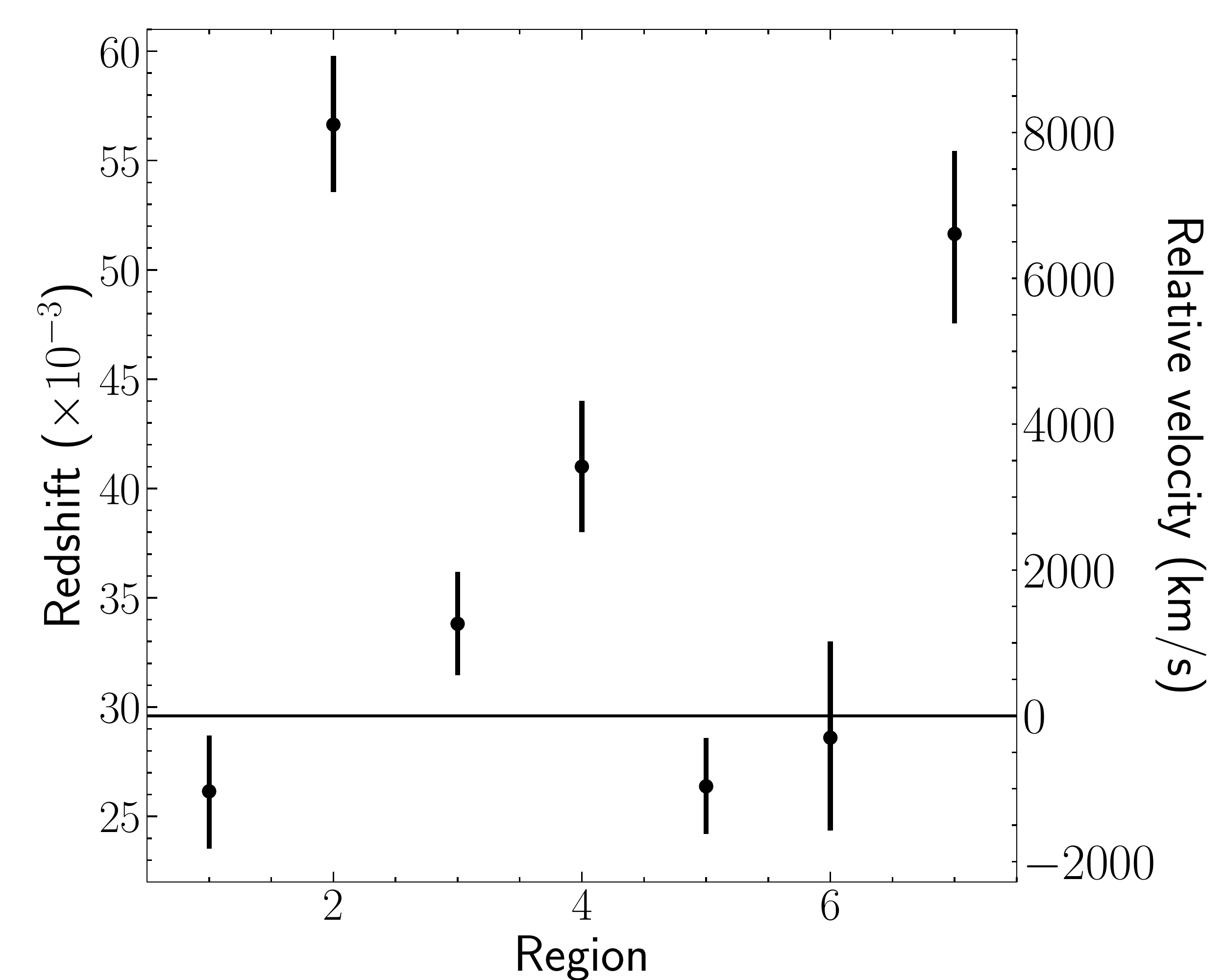}
        \caption{\emph{Top panel:} Manually selected regions following the velocity spectral map. \emph{Bottom panel:} velocities obtained for each region.  } \label{fig_ellipse_manual} 
\end{figure}

We manually analyzed selected regions close to the cluster core to analyze the redshift-blueshift pattern identified above. The top panel in Figure~\ref{fig_ellipse_manual} shows the regions analyzed while the bottom panel shows the best-fit velocities. For regions 2 and 7 the velocities correspond to $8108\pm 925$ km/s and $6609\pm 1225$ km/s, respectively. This corresponds to a departure from systemic at $> 5\sigma$ confidence level. For other regions 2-5 the velocities depart from systemic by $\sim 1-2$ $\sigma$, except for region 6 which seems to be consistent with the systemic velocity.  That is, we have found large difference in the gas velocities near the cluster core. 

\begin{figure} 
\centering 
\includegraphics[width=0.40\textwidth]{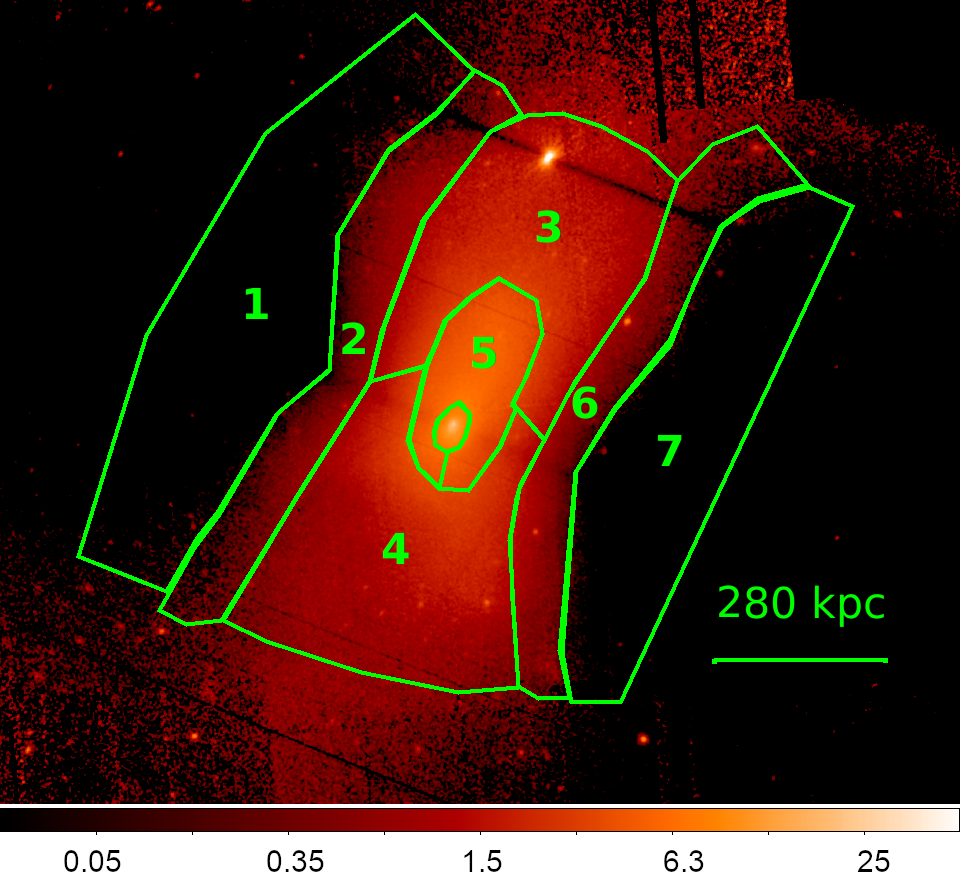}\\
\includegraphics[width=0.45\textwidth]{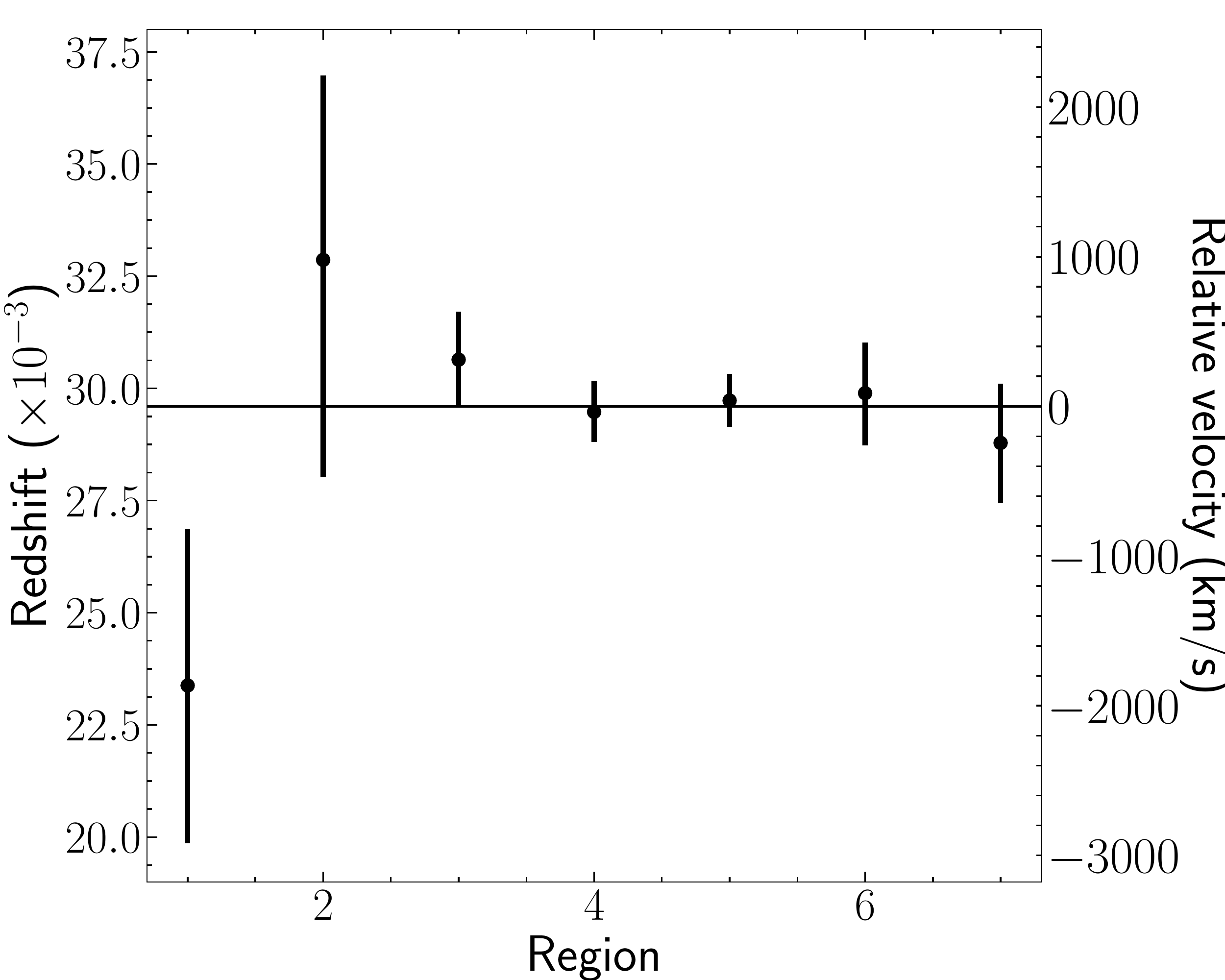}\\
\includegraphics[width=0.40\textwidth]{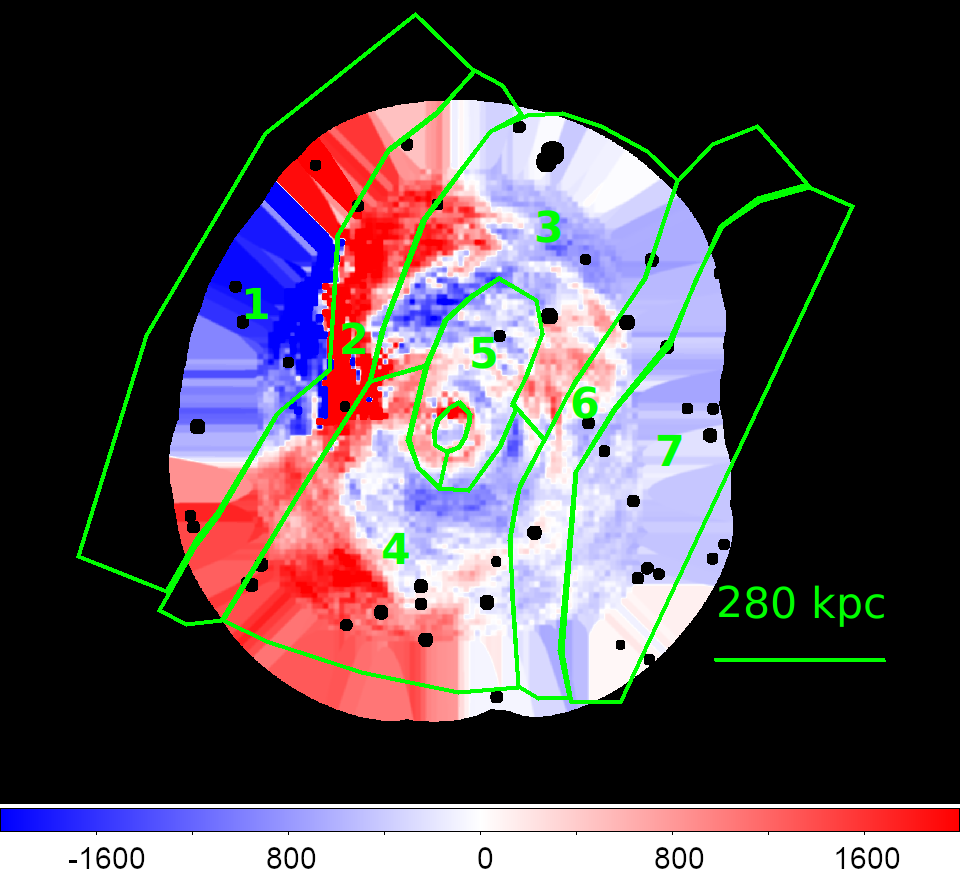}
        \caption{\emph{Top panel:} Ophiuchus cluster extracted regions for the regions following the fractional difference in the surface brightness. \emph{Middle panel:} velocities obtained for each region. The Ophiuchus redshift is indicated with an horizontal line (See Section~\ref{regions_cold_fronts}). \emph{Bottom panel:} regions analyzed superposed to the velocity spectral map obtaine in Section~\ref{spec_maps}.  } \label{fig_cold_fronts2a} 
\end{figure}

\begin{figure} 
\centering 
\includegraphics[width=0.47\textwidth]{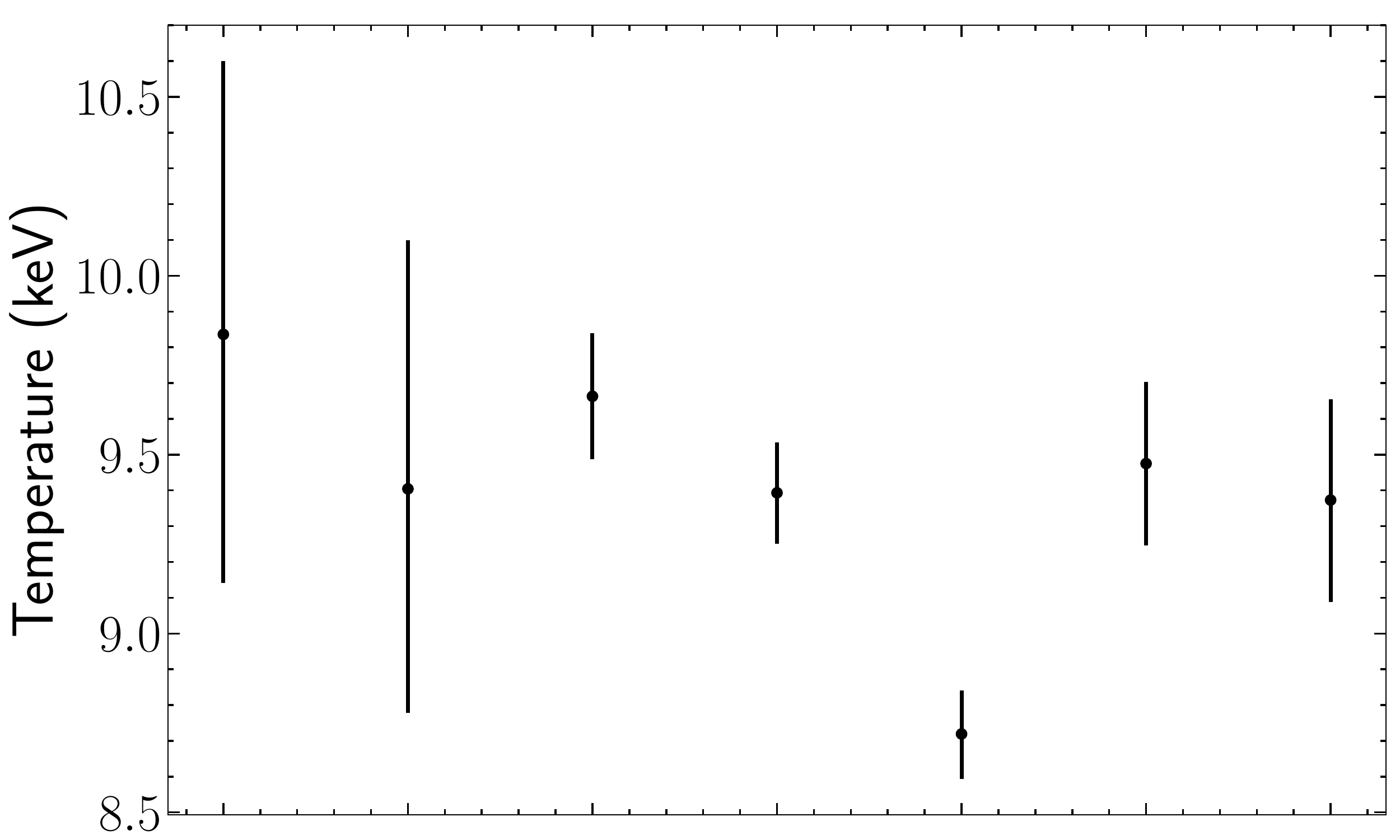}\\ 
\includegraphics[width=0.47\textwidth]{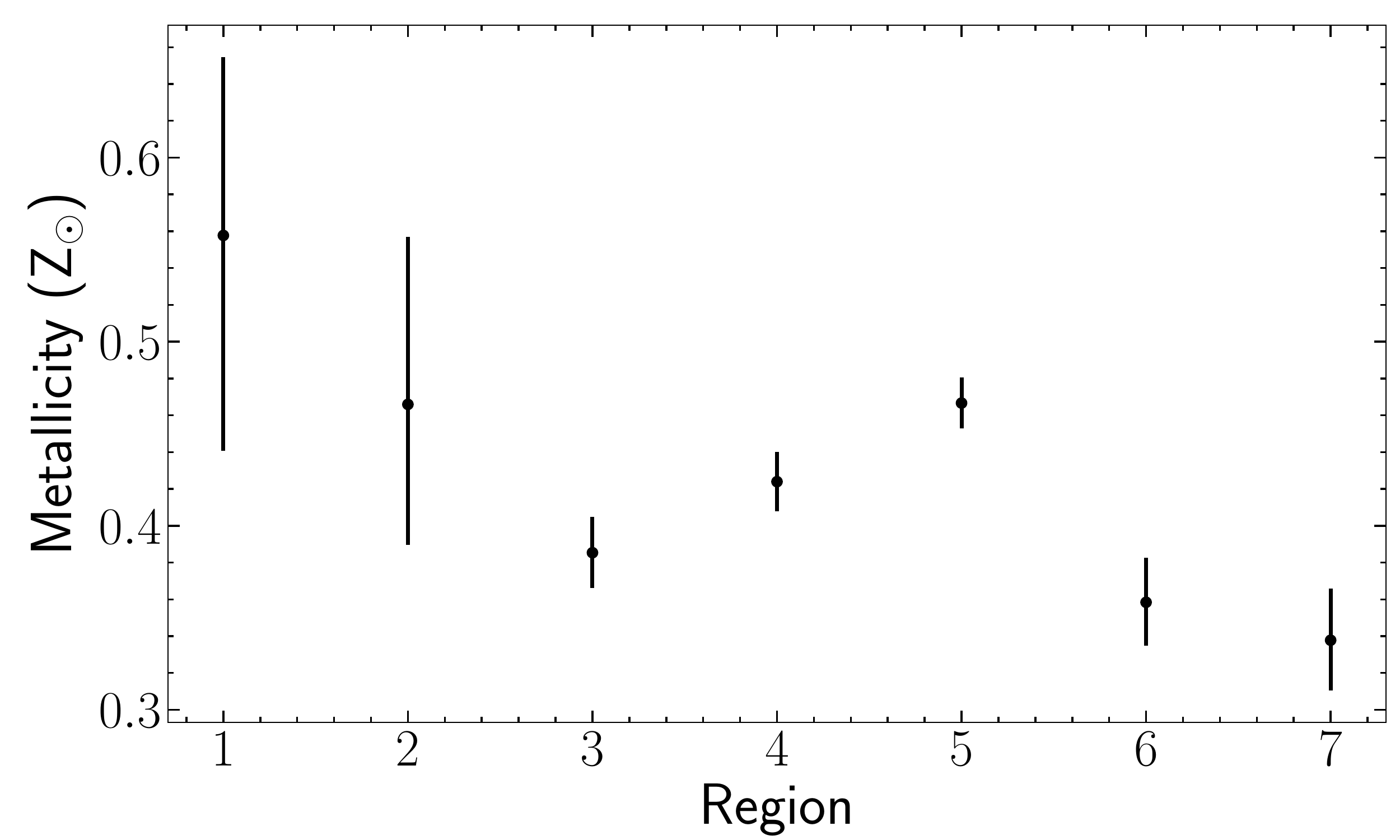}
        \caption{\emph{Top panel:} best-fit temperatures obtained for regions following the fractional difference in the surface brightness. \emph{Bottom panel:} best-fit metallicities obtained (See Section~\ref{regions_cold_fronts}). } \label{fig_cold_fronts2b} 
\end{figure}

\begin{figure*} 
\centering 
\includegraphics[width=0.85\textwidth]{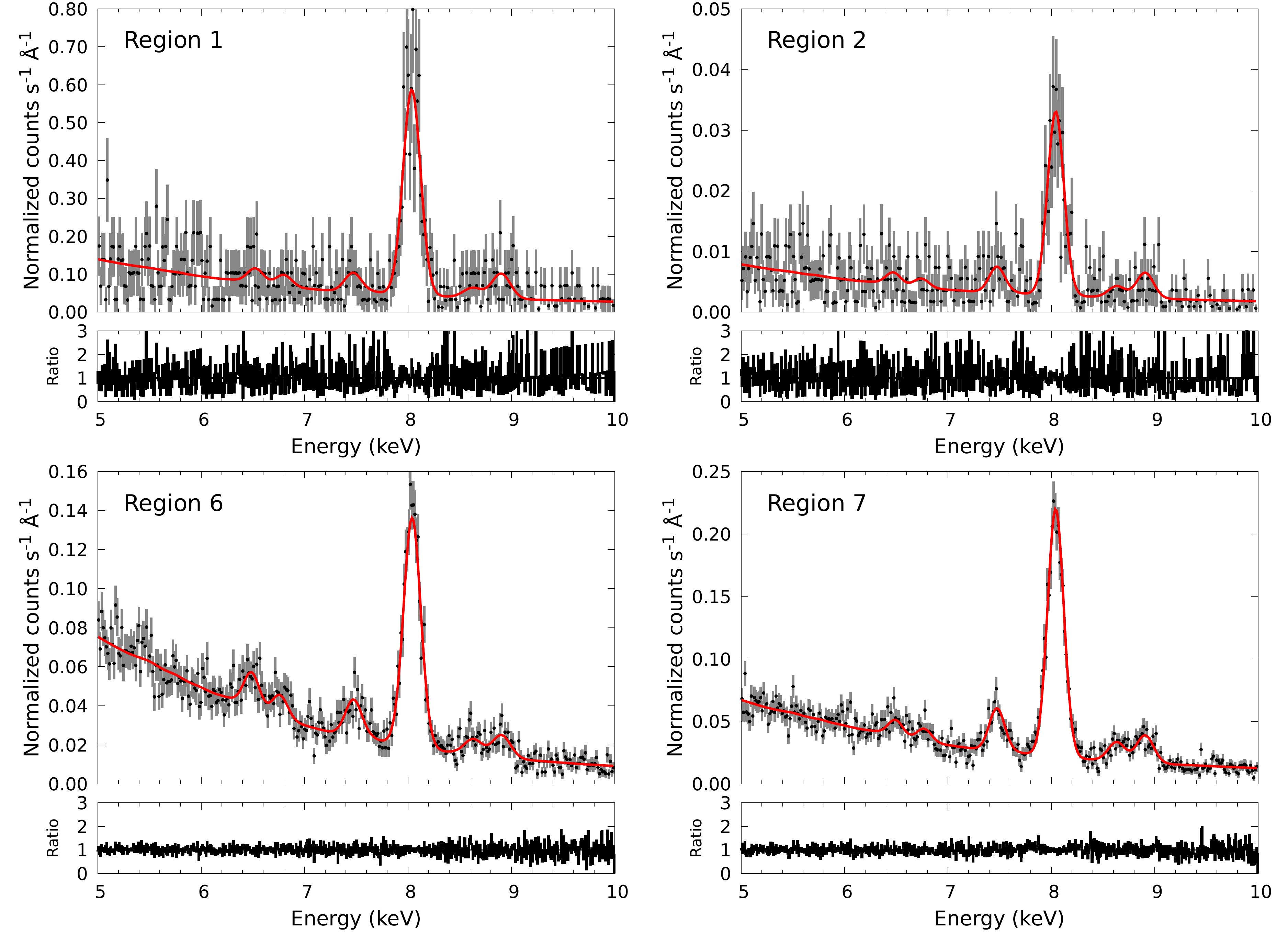}
        \caption{Best-fit spectra obtained for regions 1,2,6 and 7, as described in Section~\ref{regions_cold_fronts}.  } \label{fig_cold_fronts2a_spectra} 
\end{figure*}

\begin{table}
\scriptsize 
\caption{\label{tab_coldfront2}Ophiuchus cluster best-fit parameters for regions following the fractional difference in surface brightness. }
\centering
\begin{tabular}{ccccccc}
\\
Region &\multicolumn{5}{c}{{\tt apec} model}  \\
\hline
 &$kT$ & Z& $z$   & $norm$   & cstat/dof\\ 
  & & &    ($\times 10^{-3}$) &   ($\times 10^{-3}$)  \\ 
1&$8.89_{-1.03}^{+1.07}$&$0.51\pm 0.14$&$24.51_{-4.06}^{+4.28}$&$96.75\pm 2.08$&$1527/1570$\\
2&$8.76_{-0.80}^{+1.12}$&$0.44\pm 0.13$&$33.06_{-5.60}^{+4.86}$&$23.38\pm 0.51$&$1566/1574$\\
3&$9.66\pm 0.18$&$0.39\pm 0.02$&$30.64\pm 1.07$&$64.84\pm 0.23$&$1970/1699$\\
4&$9.39\pm 0.14$&$0.42\pm 0.02$&$29.48\pm 0.68$&$69.75\pm 0.20$&$1984/1699$\\
5&$8.72\pm 0.13$&$0.47\pm 0.01$&$29.73\pm 0.59$&$67.31\pm 0.17$&$1881/1699$\\
6&$9.48\pm 0.23$&$0.36\pm 0.02$&$29.89_{-1.17}^{+1.13}$&$18.52\pm 0.08$&$1752/1699$\\
7&$9.37\pm 0.28$&$0.34\pm 0.03$&$28.78\pm 1.34$&$14.52\mp 0.07$&$1867/1699$\\
\\ 
 \hline
\end{tabular}
\end{table}

Figure~\ref{fig_velocity_ellipses1} shows the temperature and metallicity maps (bottom panels). The ICM temperature across the cluster is generally high, with $kT>9$ in most of the regions. There is a clear drop in temperature in the cluster core, which also correspond to a large metallicity. We have found a region displaying higher temperature along the east direction from the cluster center (i.e. the conic shape), which corresponds to a high metallicity. In general, the ICM appears remarkably isothermal, similar to the results obtained by \citet{wer16}. The blueshifted-redshifted interface discussed before and located in the along the north-east direction corresponds to a region with large temperatures and low metallicities.

Using these spectral maps we calculated projected pseudo-density, pressure and entropy in each spatial bin and assuming a constant line-of-sight depth for all spectral regions. The pseudo-density was computed as $n\equiv \sqrt{\eta}$, where $\eta$ is the normalization of the {\tt apec} model. Then, we estimated pseudo-pressure as $P\equiv n\times kT$ and pseudo-entropy as $S\equiv n^{-2/3}\times kT$ \citep[see][]{hof16}. Figure~\ref{fig_entropy_ellipses} shows the density, pressure and entropy maps created from the best-fit results for the inner part of the cluster. As indicated by \citet{wer16} the pressure distribution is azimuthally asymmetric around the cluster core and displays a region of excess in the north. Such perturbed gravitational potential may be the results of a merging event with an infalling subcluster.

\subsection{Discontinuities in the surface-brightness}\label{regions_cold_fronts} 

We have manually selected multiple regions following the fractional difference in 0.5 to 9.25 keV surface brightness in order to study the velocity structure. The top panel of Figure~\ref{fig_cold_fronts2a} shows the regions used for the analysis. We have obtained velocity measurements with uncertainties down to $\Delta v\sim  170$ km/s (for region 5). Interestingly, we found a rapid change in the velocities between regions 1 ($-1596_{-1216}^{+1283}$~km/s) and 2 ($1037_{-1679}^{+1457}$~km/s), although the uncertainties are large, as shown in the spectral map (Figure~\ref{fig_velocity_ellipses1}), therefore no physical information can be obtained from such values. When excluding regions with significance $<3\sigma$ (i.e. from the Figure~\ref{fig_vel_velerror_ratio}), the difference of the uncertainties are lower ($-1864_{-742}^{+1354}$~km/s for region 1 and $978_{-1069}^{+1415}$~km/s for region 2, see Figure~\ref{fig_cold_fronts2a} middle panel). Bottom panel in Figure~\ref{fig_cold_fronts2a} shows the regions analyzed on top of the velocity map, for illustrative purposes. We noted that the regions for which the velocity suddenly changes correspond to the interface found in the velocity map. \citet{wer16} concluded that the sharp surface brightness discontinuities observed away from the core are most likely due to gas dynamics associated with a merger instead of an extraordinary AGN outburst.

Temperatures and metallicities are shown in Figure~\ref{fig_cold_fronts2b} and Table~\ref{tab_coldfront2}. We noted that temperatures and metallicities are similar between adjacent regions thus indicating a significant metal mixing at the interfaces, in contrast with the results obtained for the Virgo and Centaurus cluster by \citep{gat22a,gat22b}. Figure~\ref{fig_cold_fronts2a_spectra} shows the best fit spectra obtained for regions 1,2,6 and 7.

 \subsection{Giant Radio Fossil}\label{regions_radio_fossil}
 
\begin{figure} 
\centering 
\includegraphics[width=0.42\textwidth]{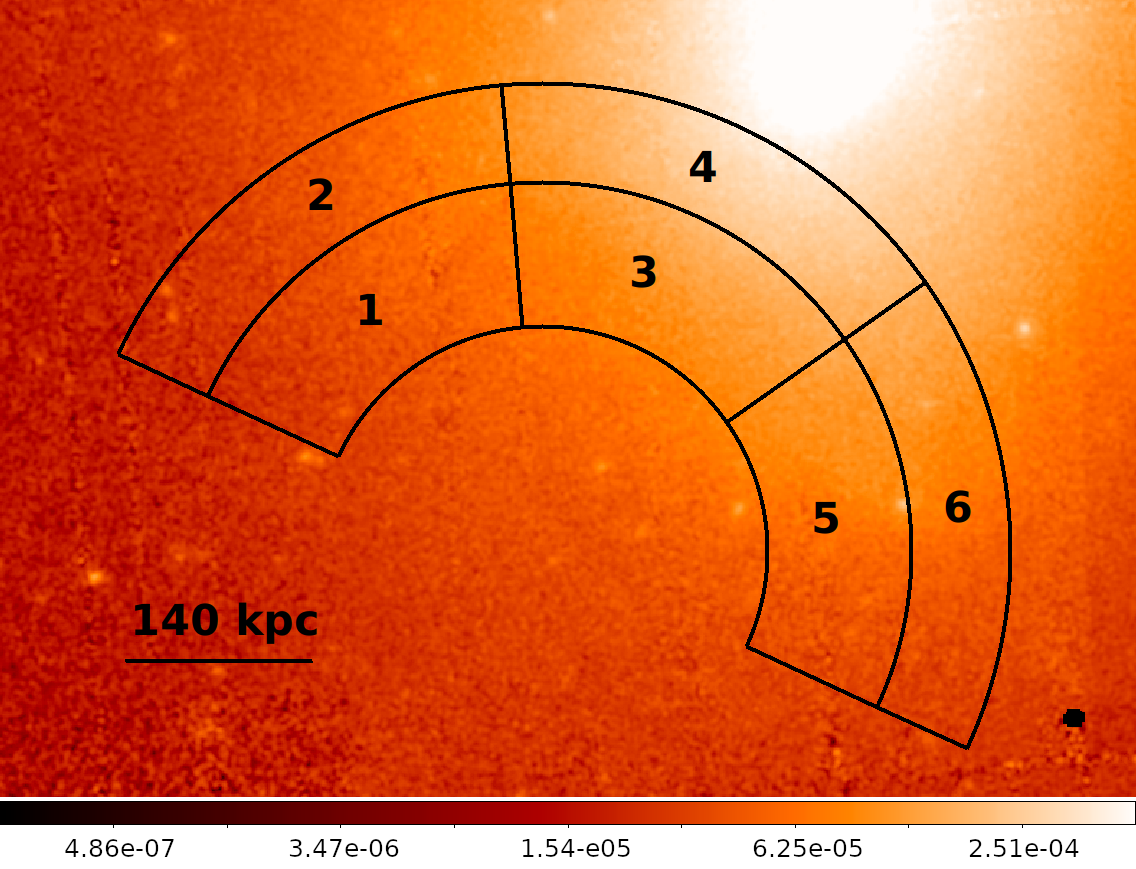}\\
\includegraphics[width=0.42\textwidth]{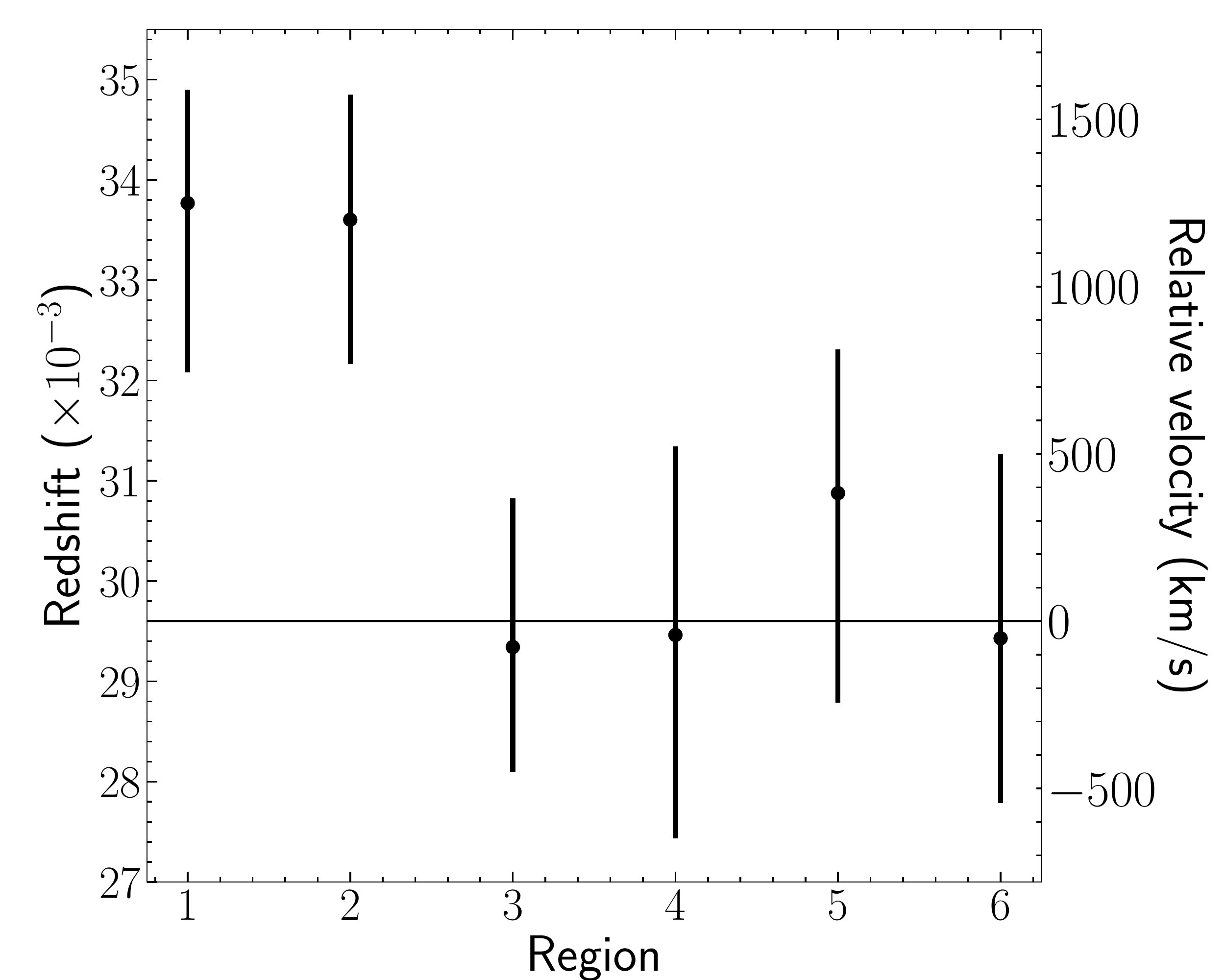}
        \caption{\emph{Top panel:} Ophiuchus cluster extracted regions for regions inside (1,3,5) and outside (2,4,6) of the giant radio fossil identified by \citet{gia20}. \emph{Bottom panel:} best-fit velocities obtained. The Ophiuchus redshift is indicated with an horizontal line (See Section~\ref{regions_radio_fossil}).  } \label{fig_radio_fossil1} 
\end{figure}

\begin{figure} 
\centering 
\includegraphics[width=0.42\textwidth]{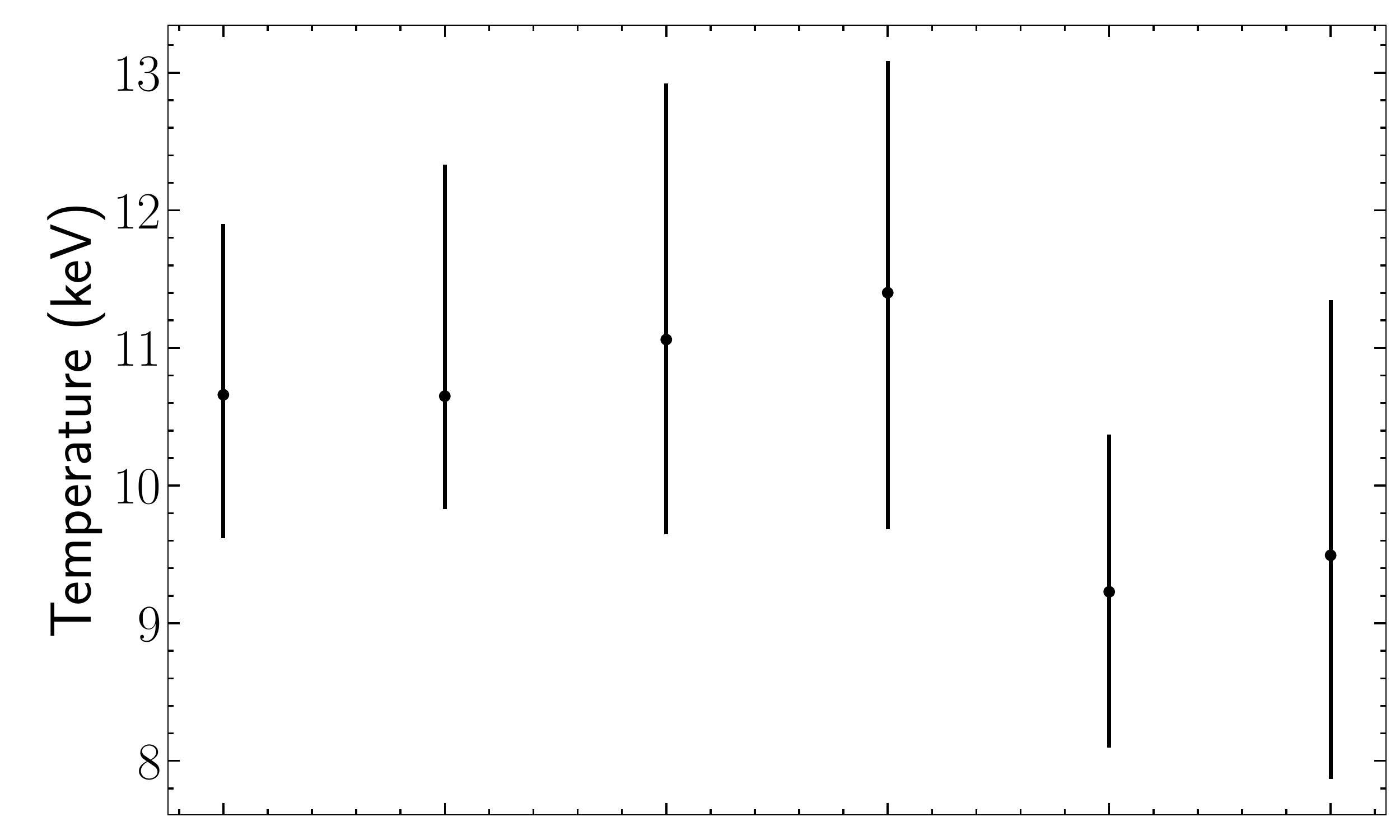}\\
\includegraphics[width=0.42\textwidth]{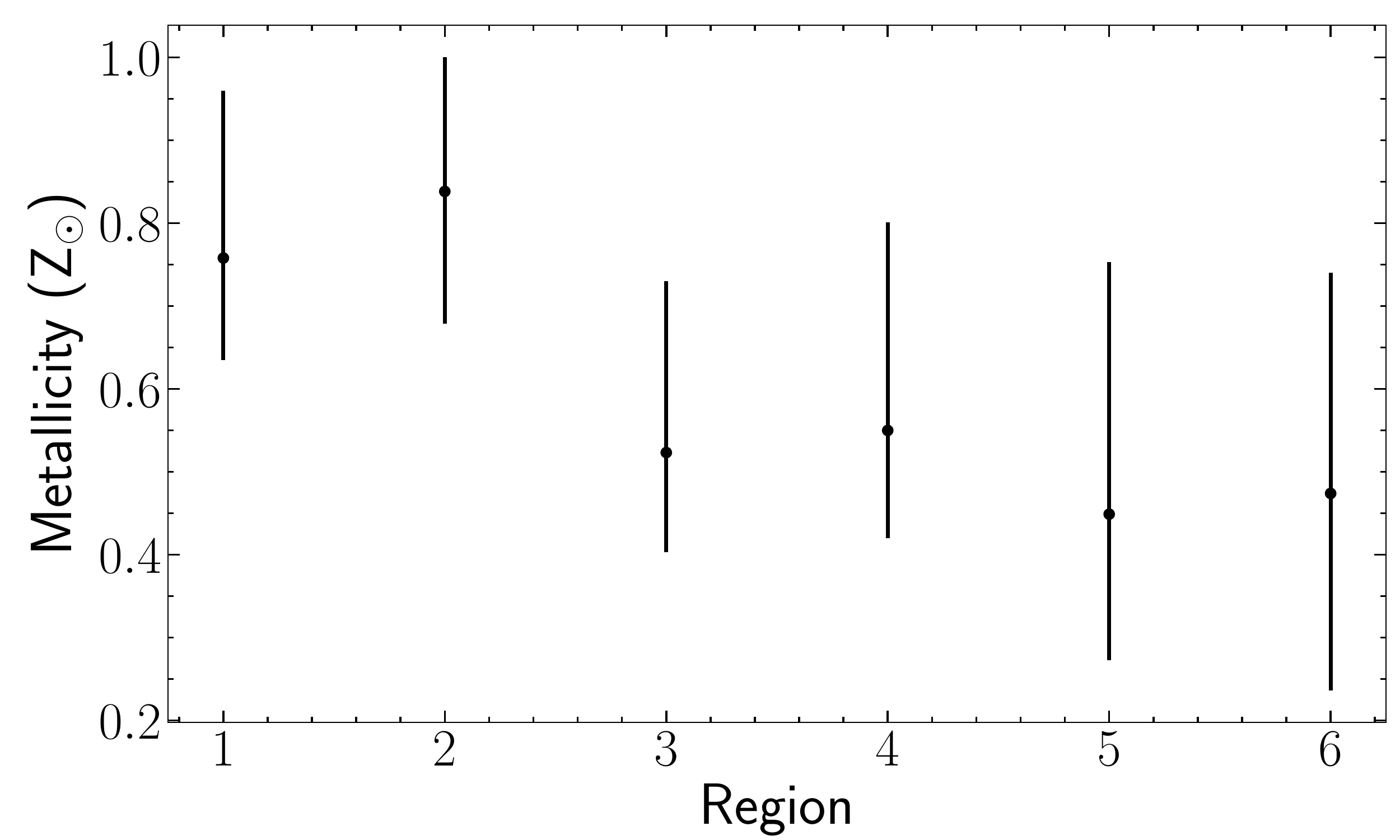}
        \caption{ \emph{Top panel:} best-fit temperatures obtained for regions inside (1,3,5) and outside (2,4,6) of the giant radio fossil identified by \citet{gia20}. \emph{Bottom panel:} best-fit metallicities obtained (See Section~\ref{regions_radio_fossil}). } \label{fig_radio_fossil2} 
\end{figure}

\citet{gia20} identified a giant cavity in the X-ray gas filled with diffuse radio emission with an extraordinarily steep radio spectrum. Such structure may be an aged fossil of a very powerful AGN outburst ($pV\sim 5\times 10^{61}$ erg for this cavity). We analyzed the X-ray spectra located inside and outside the structure. The top panel of Figure~\ref{fig_radio_fossil1} shows the semi-annular regions analyzed. Following \citet{gia20}, the inner semi-annular region (i.e. regions 1, 3 and 5) correspond to a circular region of radius $r=6^{\prime}.8$, centered on RA$_{\rm J2000}$=17h12m50.2s and DEC$_{\rm J2000}$=$-23$d31m13s. The lower panel of Figure~\ref{fig_radio_fossil1} shows the velocities obtained. The error-weighted standard deviation of the data points is $\sigma=552$~km/s. The velocities obtained for regions 1 and 2 are slightly larger than those obtained for the other regions ($\sim2.3\sigma$ confidence level). Figure~\ref{fig_radio_fossil2} shows the best-fit temperatures (top panel) and metallicities (bottom panel) obtained. We found that temperatures and metallicities are similar between the different regions analyzed. Such feature could be a hint of gas sloshing \citep{roe11,gat22a,gat22b}, although uncertainties are large.

\subsection{Energy budget}\label{sec_dis}  
\begin{figure} 
\centering 
\includegraphics[width=0.40\textwidth]{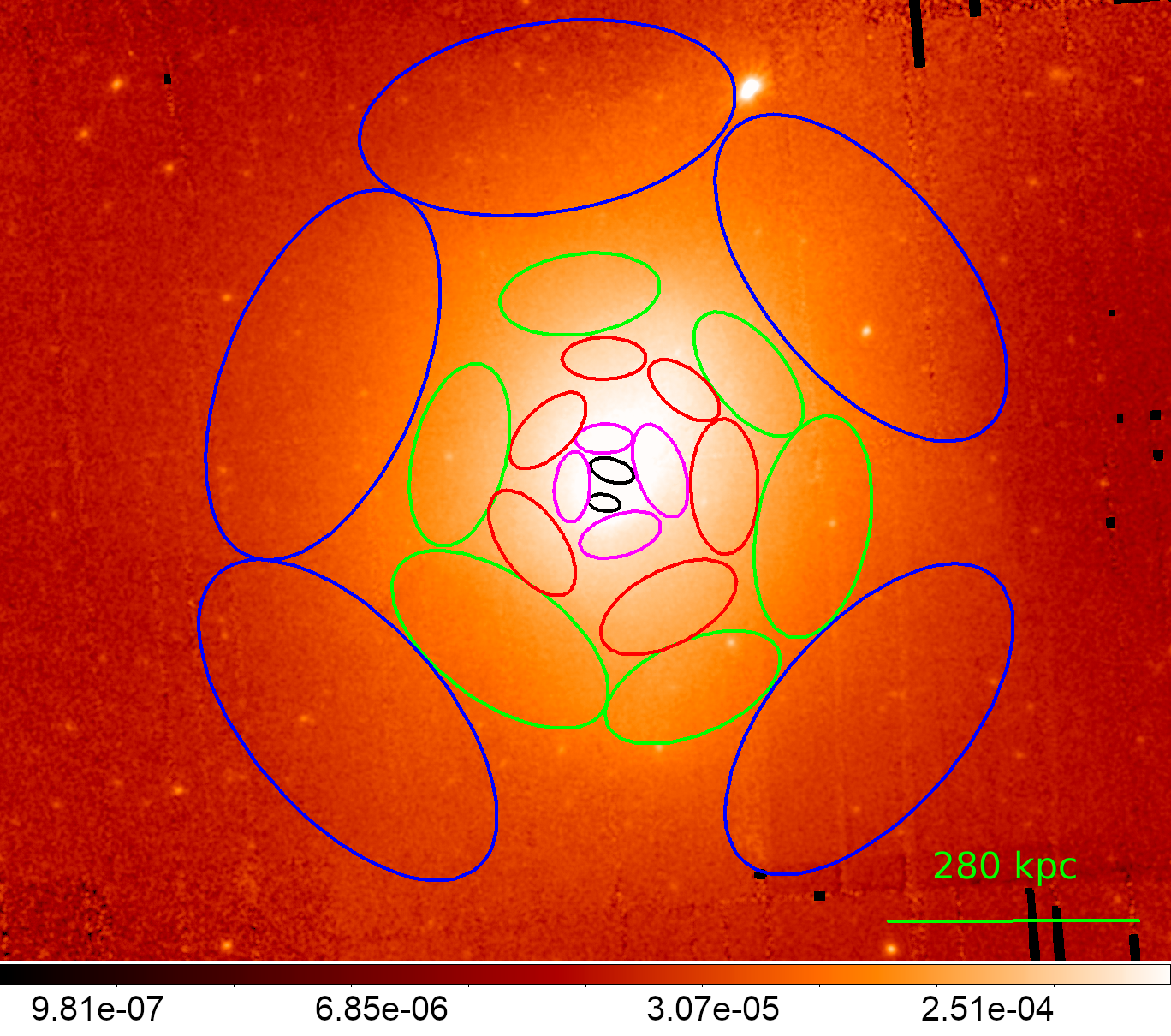}
\caption{Non-overlapping regions selected to analyze the energy profile for the spectral map obtained in Section~\ref{spec_maps}. Different colors indicate the different radii assigned.} \label{fig_energies1} 
\end{figure}

\begin{figure} 
\centering 
\includegraphics[width=0.40\textwidth]{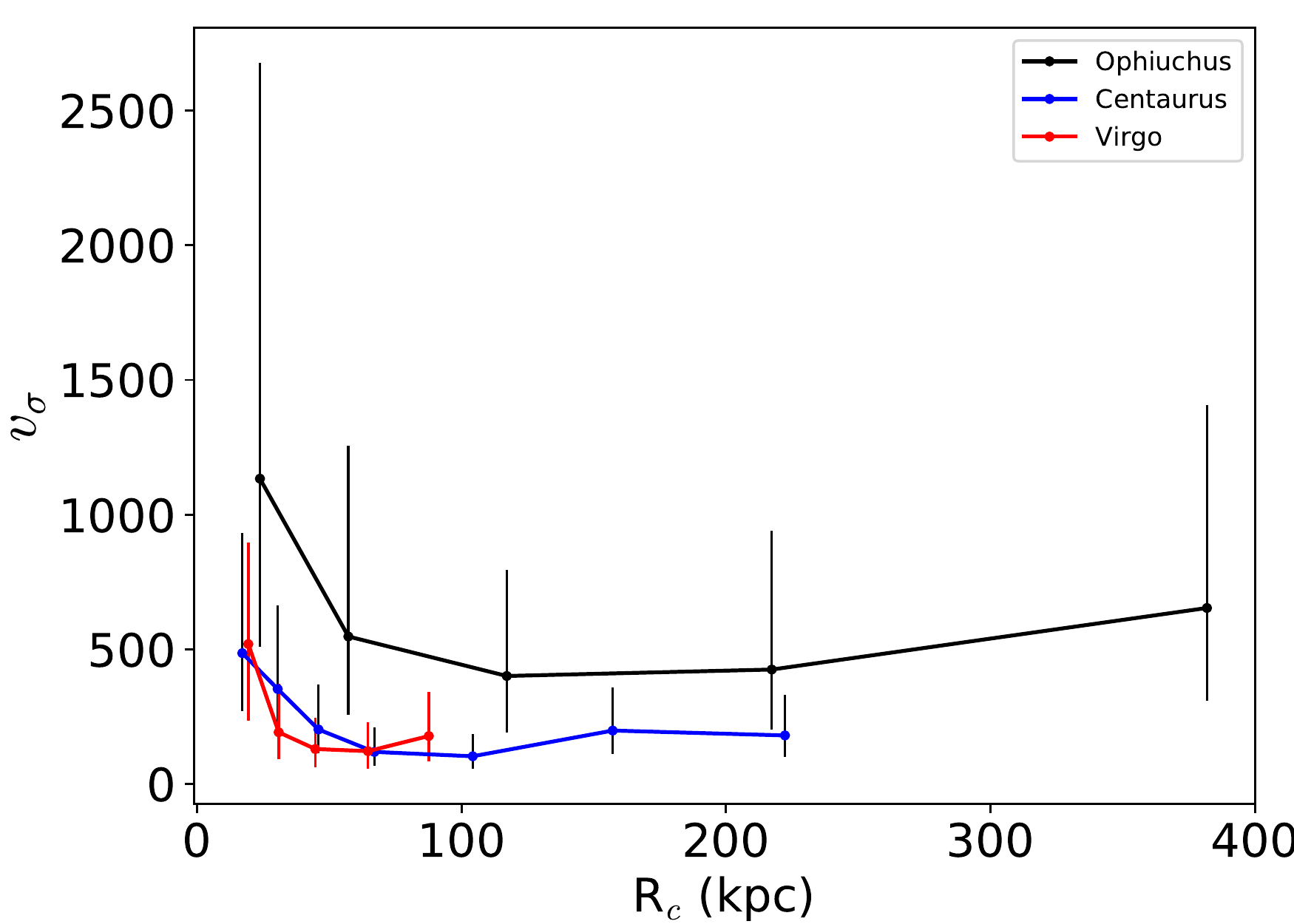}\\
\includegraphics[width=0.40\textwidth]{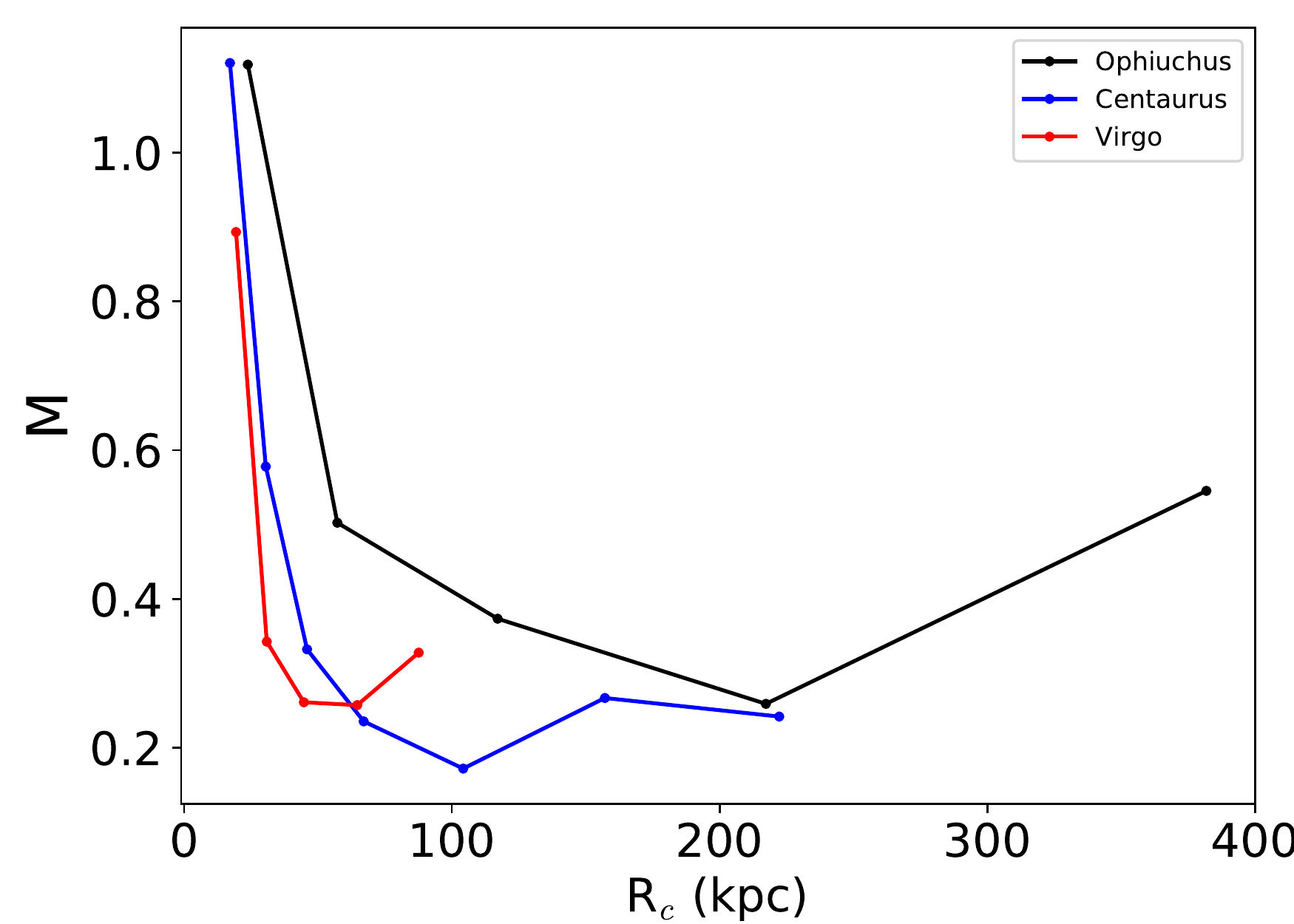} \\
\includegraphics[width=0.40\textwidth]{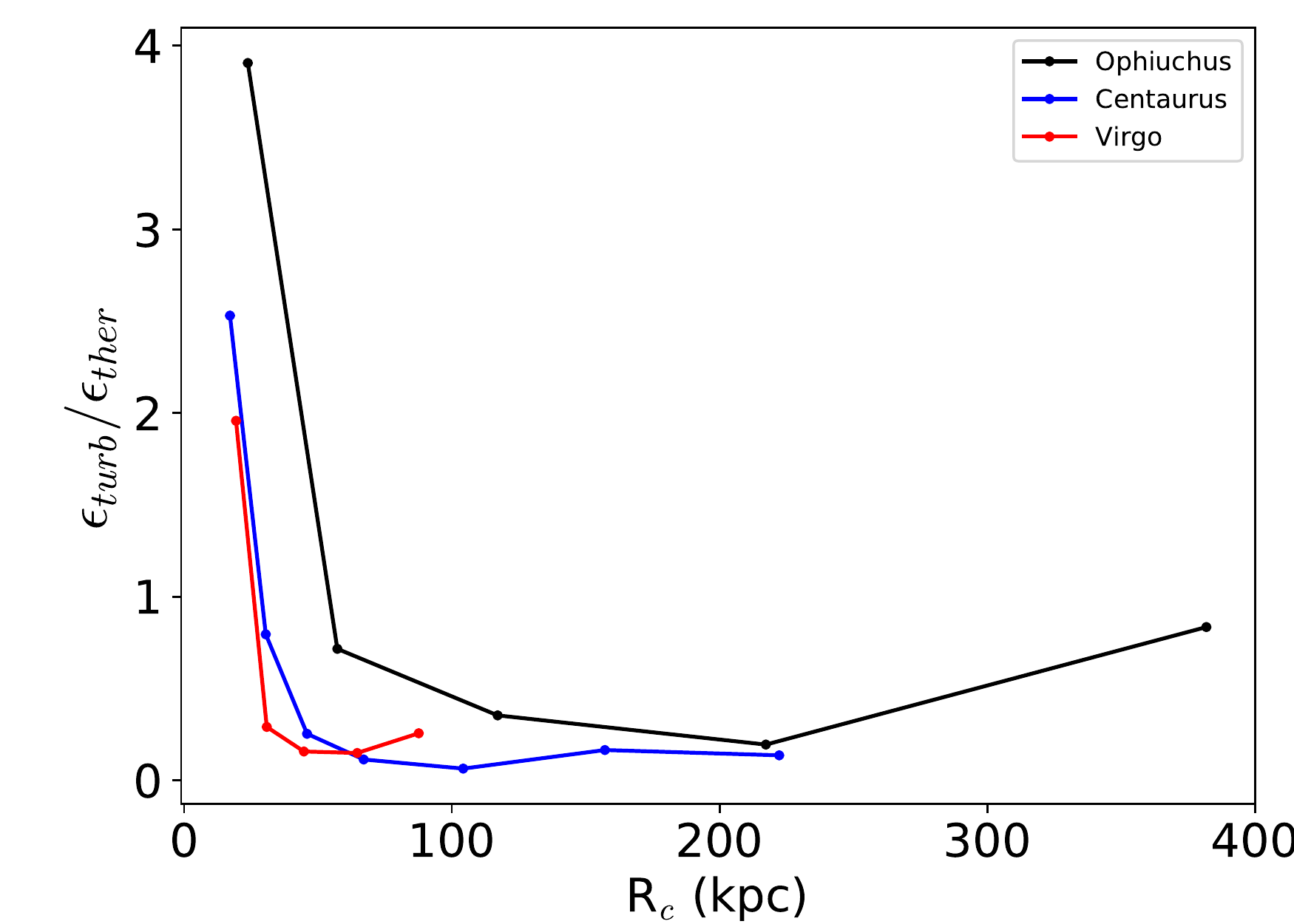} 
\caption{\emph{Top panel:} width of the velocity distribution as function of the distance from the cluster center. \emph{Middle panel:} mach number as function of the distance from the cluster center. \emph{Bottom panel:} upper limits for the ratio of turbulent to thermal energy ($\epsilon_{tur}/\epsilon_{ther}$). For comparison, values obtained for the Virgo and Centaurus clusters by \citet{gat22a,gat22b} are also included. } \label{fig_energies2} 
\end{figure}

For different radii we measured velocities for non-overlapping elliptical regions obtained in the spectral maps (see Section~\ref{spec_maps}). 
Figure~\ref{fig_energies1} shows the regions analyzed.
Different colors indicate the different radii assigned. 
Assuming a Gaussian distribution, we measured the $\sigma$-width and velocity mean for each radius as a function of the distance to the cluster center. 
This is done by computing a Gaussian log likelihood, adding the model width to the measurement uncertainty in quadrature to produce the Gaussian width (i.e. $\sigma$).  
A Markov chain Monte Carlo (MCMC) method is then used to sample the likelihood. 
The median chain values produce the velocity width, center and the 1$-\sigma$ percentiles. 
The top panel of Figure~\ref{fig_energies2} shows the $\sigma$-width distribution as a function of the distance to the cluster center. 
As a comparison, we included the values obtained for the Virgo and Centaurus clusters by \citet{gat22a,gat22b}. 
There is a hint from the plot that the $\sigma$-width decreases as we move away from the inner radius although large uncertainties prevent us from making firmer conclusions.

We performed simulations in order to demonstrate that this method provides reasonable constraints on the $\sigma$-width and velocity mean for each radius. First, we simulate a sample of $N=10000$ velocities including their uncertainties and following a Gaussian distribution. Then, we applied the MCMC method to estimate the already know $v_{\sigma}$ value. We have found that including only two random points from the sample (i.e. the inner region in Figure~\ref{fig_energies1}) the $v_{\sigma}$ can be overestimated up to $1.5$ times the real value. Taking 4 points we can estimate $v_{\sigma}$ to $25\%$ the real value. Therefore, the values obtained for the innermost region in the Ophiuchus cluster analysis should be taken with caution as they may be overestimated.

We compute the sound speed for each region as $c_{s}=\sqrt{\gamma kT/\mu m_{p}}$, where $\gamma$ is the adiabatic index (5/3), $\mu$ is the mean particle mass, $m_{p}$ is the proton mass and $kT$ is the best-fit temperature. 
Then, we computed the Mach number ($M=v_{turb}/c_{s}$) as function of the radius (see middle panel in Figure~\ref{fig_energies2}). 
We have found that for the innermost radius the Mach number is $1.12\pm 0.78$, similar to the values obtained for the Centaurus cluster. 
As we increase the radius the Mach number decreases, however the values are larger than those obtained for Virgo and Centaurus clusters. 
However, it is important to note that the uncertainties obtained for Ophiuchus are much larger than those obtained in our previous work.
Bottom panel in Figure~\ref{fig_energies2} shows upper limits for the ratio of turbulent ($\epsilon_{turb}$) to thermal ($\epsilon_{ther}$) energy computed as $\epsilon_{turb}/\epsilon_{ther}=\frac{\gamma}{2}M^{2}$.  
These upper limits have been computed by doing standard error propagation. 
Thus, while the Mach number for the innermost part is similar to that obtained for Centaurus \citep{gat22b}, the upper limits for the energy ratio are significantly larger. 
We found a contribution from the turbulent component $<25\%$ for large radius.

\section{Discussion}\label{sec_dis} 

\citet{wer16} found multiple substructure near the Ophiuchus cluster center by analyzing {\it Chandra} observations, including a sharp eastern inner cold front with a tongue-like extension to the north, a possible Kelvin-Helmholtz roll and a western discontinuity. We are not able to study the velocity distribution in such structures due to the lower spatial resolution in the {\it XMM-Newton} observations analyzed. By analyzing the amplitude of density fluctuations of the cooling core, \citet{wer16} estimate the velocities to be lower than $\sim 100$ km/s for distances $<100$~kpc. However, they also indicated that the velocities may be underestimated due to the damping of density fluctuations by thermal conduction in the hot ICM.  The large velocities obtained in our analysis, on the other hand, point out the strong dynamic activity in the cluster. The galaxy velocity dispersion of the cluster is $\sim 954\pm 58$ km/s \citep{dur15}. However, due to the difference in the origin of the velocity distribution structure, whether turbulence or bulk motions, we expect the results to be not exactly the same. \citet{fuj08} measure upper limits in the radial velocity difference between different regions of the ICM to be less than $\sim 3000$~km/s by measuring the redshifts of metal lines in X-ray {\it Suzaku} spectra, in good agreement with the difference measured in our analysis. While these results provide constraints on velocities, future observations with the next generation of X-ray observatories (e.g. {\it Athena}, {\it XRISM}) will provide better measurements of the ICM velocity structure in this cluster and can be used to better calibrate the scale and the method for the existing (or future CCD measurements).

\section{Conclusions and summary}\label{sec_con}  

We have analyzed the velocity structure in the Ophiuchus cluster using the technique developed by \citet{san20,gat22a,gat22b} to calibrate the absolute energy scale of the {\it XMM-Newton} EPIC-pn detector. In this Section we briefly summarize our findings.

\begin{enumerate}
\item  We have studied the velocity distribution by creating non-overlapping circular regions. Our results are consistent with a large change in velocities from blueshifted to redshifted gas located at $\sim 250$ kpc.

\item We have analyzed the velocity distribution along W and E directions, by creating non-overlapping circular regions. We have found that for distances $\sim 250$~kpc the velocities obtained for both directions tend to be opposite. 

\item  We have created 2D projected maps for velocity, temperature, metallicity, density, entropy and pressure distribution for the Ophiuchus cluster. We have found a large redshifted-blueshifted interface located $\sim 150$~kpc in the E direction from the cluster core. Our modeling of regions following the surface brigthness indicate that this interface is traced by discontinuities on it with a velocity difference of $\Delta v\sim 2500$ km/s. However, such structure in the velocities is not found along a similar surface brightness discontinuity located in the W direction from the cluster core. The interface displays hints for lack of metal mixing. We have obtained accurate velocity measurements with uncertainties down to $\Delta v \sim 170$ km/s. We found large velocity differences between some regions near the cluster core, with departure from systematics $>5\sigma$.

\item We have found that the velocities near the cluster core are similar to that from the main system.

\item We found that metallicities and temperatures are similar inside/outside the radio fossil identified by \citet{gia20} while there are hints for change in the velocities. 

\item We have computed the the width of the velocity distribution decreases as we move away from the cluster center. We have found a contribution from the turbulent component of $<25\%$ to the total energy budget at different radii.
\end{enumerate} 
 
Future work will include a detailed analysis of the multitemperature component as well as elemental abundances spatial distribution by including the soft energy band ($<1.5$~keV).  

\section{Acknowledgements} 
This work was supported by the Deutsche Zentrum f\"ur Luft- und Raumfahrt (DLR) under the Verbundforschung programme (Messung von Schwapp-, Verschmelzungs- und R\"uckkopplungsgeschwindigkeiten in Galaxienhaufen.). This work is based on observations obtained with XMM-Newton, an ESA science mission with instruments and contributions directly funded by ESA Member States and NASA. This research was carried out on the High Performance Computing resources of the cobra cluster at the Max Planck Computing and Data Facility (MPCDF) in Garching operated by the Max Planck Society (MPG)
\subsection*{Data availability}
The observations analyzed in this article are available in {\it XMM-Newton} Science Archive (XSA\footnote{\url{http://xmm.esac.esa.int/xsa/}}).

\bibliographystyle{mnras}

%

 \appendix\label{sec_apx}

 \begin{figure} 
\centering 
\includegraphics[width=0.45\textwidth]{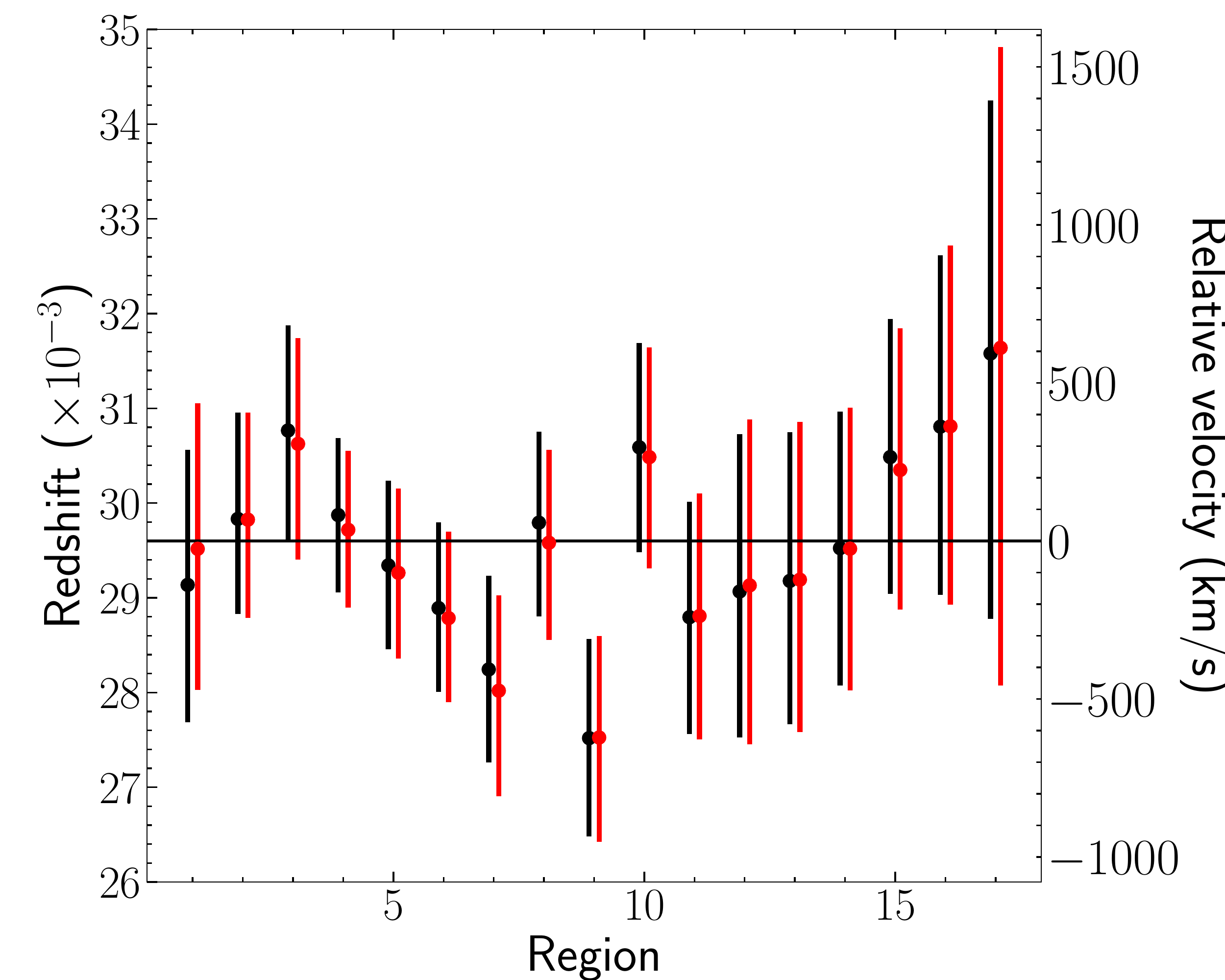}
\caption{Velocities obtained for each region for Case 1 (numbered from the center to the outside) with the {\tt apec} model (black points) and the {\tt lognorm} model (red points). The Ophiuchus redshift is indicated with a horizontal line. } \label{fig_velocity_ellipses_apec_comparison} 
\end{figure}

\section{Fitting the data with a multi-temperature model}\label{sec_single_apec}  
We tested the {\tt lognorm} multi-temperature model for the \emph{Case 1} analysis described in Section~\ref{circle_rings} in order to compare it with the {\tt apec} model results. While temperatures and metallicity obtained values are different we found similar velocities between both models (see Figure~\ref{fig_velocity_ellipses_apec_comparison}). Future work will include the soft energy band ($<1.5$~keV) and a detailed analysis of the temperature distribution within the cluster.

\end{document}